\documentclass[reprint,aps,prl,amsmath,amssymb,nofootinbib,superscriptaddress]{revtex4-2}

\usepackage[utf8]{inputenc}
\usepackage[T1]{fontenc}

\usepackage{amsmath}
\usepackage{txfonts}
\usepackage{mathtools}
\usepackage{braket}
\usepackage{tensor}
\usepackage{xcolor}
\usepackage[abbreviations]{siunitx}
\usepackage{graphicx}
\usepackage{booktabs}

\usepackage{float}

\usepackage{braket}

\newcommand{\eMax}{\ensuremath{e_{\text{Max}}}}

\usepackage{hyperref}
\usepackage{cleveref}

\newcommand{\beq}{\begin{equation}}
\newcommand{\eeq}{\end{equation}}
\newcommand{\beqn}{\begin{eqnarray}}
\newcommand{\eeqn}{\end{eqnarray}}
\newcommand{\bsub}{\begin{subequations}}
\newcommand{\esub}{\end{subequations}}
\newcommand{\bpm}{\begin{pmatrix}}
\newcommand{\epm}{\end{pmatrix}}

\DeclareSIUnit{\fm}{\femto\meter}
\DeclareSIUnit{\MeVc}{\MeV\per\text{\ensuremath{c}}}

\allowdisplaybreaks[3]

\begin{document}
\title{\emph{Ab initio} uncertainty  quantification of neutrinoless double-beta decay in $^{76}$Ge}

\author{A.~Belley}
\affiliation{TRIUMF, Vancouver, BC, Canada}
\affiliation{Department of Physics \& Astronomy, University of British Columbia, Vancouver, BC, Canada}

\author{J. M. Yao}   
\affiliation{School of Physics and Astronomy, Sun Yat-sen University, Zhuhai 519082, P.R. China}

\author{B.~Bally}
\affiliation{ESNT, IRFU, CEA, Université Paris-Saclay, 91191 Gif-sur-Yvette, France.}

\author{J.~Pitcher}
\affiliation{TRIUMF, Vancouver, BC, Canada}
\affiliation{Department of Physics \& Astronomy, University of British Columbia, Vancouver, BC, Canada}

\author{J.~Engel} 
\affiliation{Department of Physics and Astronomy, University of North Carolina, Chapel Hill, North Carolina 27516-3255, USA. }

\author{H.~Hergert} 
\affiliation{Facility for Rare Isotope Beams, Michigan State University, East Lansing, Michigan 48824-1321, USA.}
\affiliation{Department of Physics \& Astronomy, Michigan State University, East Lansing, Michigan 48824-1321, USA.}

\author{J.~D.~Holt} 
\affiliation{TRIUMF, Vancouver, BC, Canada}
\affiliation{Department of Physics, McGill University, Montr\'eal, QC, Canada.}

\author{T.~Miyagi}
\affiliation{Technische Universit\"at Darmstadt, Department of Physics, 64289 Darmstadt, Germany.}
\affiliation{ExtreMe Matter Institute EMMI, GSI Helmholtzzentrum f\"ur Schwerionenforschung GmbH, 64291 Darmstadt, Germany.}
\affiliation{Max-Planck-Institut f\"ur Kernphysik, Saupfercheckweg 1, 69117 Heidelberg, Germany}

\author{T.~R.~Rodr{\'i}guez}
\affiliation{Departamento de Estructura de la Materia, F\'isica T\'ermica y Electr\'onica, Universidad Complutense de Madrid, E-28040 Madrid, Spain.}
\affiliation{Departamento de F\'isica Te\'orica, Universidad Aut\'onoma de Madrid, E-28049 Madrid, Spain.}
\affiliation{Centro de Investigaci\'on Avanzada en F\'isica Fundamental-CIAFF-UAM, E-28049 Madrid, Spain.}

\author{A.~M.~Romero}
\affiliation{Departament de Física Quàntica i Astrofísica (FQA), Universitat de Barcelona (UB), c. Martí i Franqués, 1, 08028 Barcelona, Spain.}
\affiliation{Institut de Ciències del Cosmos (ICCUB), Universitat de Barcelona (UB), c. Martí i Franqués, 1, 08028 Barcelona, Spain.}

\author{S.~R.~Stroberg}
\affiliation{Department of Physics and Astronomy, University of Notre Dame, Notre Dame, IN 46556 USA.}

\author{X.~Zhang}   
\affiliation{School of Physics and Astronomy, Sun Yat-sen University, Zhuhai 519082, P.R. China}

\begin{abstract}
The observation of neutrinoless double-beta ($0\nu\beta\beta$) decay would offer proof of lepton number violation, demonstrating that neutrinos are Majorana particles, while also helping us understand why there is more matter than antimatter in the Universe. 
If the decay is driven by the exchange of the three known light neutrinos, a discovery would, in addition, link the observed decay rate to the neutrino mass scale through a theoretical quantity known as the nuclear matrix element (NME).  Accurate values of the NMEs for all nuclei considered for use in $0\nu\beta\beta$ experiments are therefore crucial for designing and interpreting those experiments.
Here, we report the first comprehensive ab initio uncertainty quantification of the $0\nu\beta\beta$-decay NME, in the key nucleus \nuclide[76]{Ge}.  Our method employs nuclear strong and weak interactions derived within chiral effective field theory and recently developed many-body emulators.  Our result, with a conservative treatment of uncertainty, is an NME of 
$2.60^{+1.28}_{-1.36}$, which, together with the best-existing half-life sensitivity and phase-space factor, sets an upper limit for effective neutrino mass of $187^{+205}_{-62}$ meV.  The result is important for designing next-generation germanium detectors aiming to cover the entire inverted hierarchy region of neutrino masses.  
 
\end{abstract}

\maketitle

\paragraph{\textbf{Introduction.}} 
 The origin of the matter–antimatter asymmetry in the Universe remains one of the most important unsolved puzzles in physics. 
 Many theories suggest that the asymmetry originates from a violation of lepton number through {\em leptogenesis}~\cite{Fukugita:1986PLB}, in which leptons are created with no corresponding antileptons. 
 The most promising way at present to determine the level at which lepton number is violated is through the hypothetical nuclear transition known as neutrinoless double-beta ($0\nu\beta\beta$) decay~\cite{Furry:1939}, in which two neutrons inside an atomic nucleus are transmuted into two protons, and two electrons are emitted without any of the antineutrinos that lepton-number conservation requires.
 The detection of $0\nu\beta\beta$ decay would immediately demonstrate that neutrinos are Majorana fermions~\cite{Schechter:1982PRD}, i.e., their own antiparticles, and therefore have significant implications for the Universe's matter–antimatter asymmetry.
 Furthermore, if $0\nu\beta\beta$ decay is mediated by light Majorana neutrino exchange, its half-life can be related to an effective neutrino mass $\langle m_{\beta\beta}\rangle=\sum_iU^2_{ei}m_i$, where $m_i$ are the masses of light neutrinos, and $U_{ei}$ are elements of the unitary matrix that mixes electron neutrinos with other flavors.  The precision with which $\langle m_{\beta\beta}\rangle$ can be determined depends on how well the nuclear matrix element (NME) that governs the decay can be calculated.   
  
\nuclide[76]{Ge} is one of only a few highly promising candidate nuclei for experiments, as germanium detectors possess the advantages of high energy resolution, low internal background, and high detection efficiency. Several experiments  have been searching for $0\nu\beta\beta$ decay in this isotope, including the GERDA \cite{GERDA:Agostini-PRL2020}, Majorana Demonstrator (MJD)~\cite{Majorana:2022}, and CDEX \cite{CDEX:2023} collaborations. The highest half-life sensitivity so far has been reported by the GERDA experiment, which set a limit $T_{1/2} > 1.8 \times 10^{26}$ years at 90\% confidence level (C.L.) \cite{GERDA:Agostini-PRL2020}.  If light-neutrino exchange is responsible, this half-life limit establishes an upper limit for the effective neutrino mass of $\langle m_{\beta\beta}\rangle=73-204$ meV. The large range is due mainly to a spread of about a factor of three in the NMEs predicted by different nuclear models~\cite{Menendez:2009,Rodriguez:2010,Barea:2015,Horoi:2016,Song:2017,Fang:2018,Coraggio:2020}.  The spread can be even larger when the NMEs from all existing calculations with different parametrizations are considered.  The associated uncertainty is difficult to reduce because each model has its particular phenomenological assumptions and uncontrolled approximations~\cite{Engel:2015,Engel:2017,Yao:2022PPNP,Agostini:2022Review}. 

In recent years, significant progress has been made in calculating NMEs from first principles.  The required advances in {\em ab initio} nuclear theory have followed the parallel development of nuclear forces from chiral effective field theory ($\chi$EFT)~\cite{Epelbaum:2009RMP,Machleidt:2011PR}, a systematically improvable low-energy expansion of QCD, where undetermined low-energy constants (LECs) are optimized to data in few-nucleon systems, and similarity-renormalization-group (SRG) methods~\cite{Bogner:2010} for evolving such forces to the low-energy scale typical for atomic nuclei.
With the resulting interactions and operators, the $A$-body Schr\"odinger equation can now be solved fairly accurately for most atomic nuclei in the medium-mass region~\cite{Stroberg:2021PRL}, and even in the \nuclide[208]{Pb} region~\cite{Hu:2022}, by employing nonperturbative and systematically improvable many-body methods. The application of {\em ab initio} methods to $0\nu\beta\beta$ decay is important because theoretical uncertainties related to the many-body wave functions and transition operators become controllable. 

So far, three \emph{ab initio} methods, the in-medium generator coordinate method (IM-GCM)~\cite{Yao:2020PRL}, the valence-space formulation of the in-medium SRG (VS-IMSRG)~\cite{Belley:2021PRL}, and coupled-cluster theory~\cite{Novario:2021PRL}, have been used to calculate the NME of $^{48}$Ca, the lightest nucleus that could be used in an experiment. 
When starting from the same chiral two-nucleon-plus-three-nucleon (NN+3N) interaction and $0\nu\beta\beta$-decay operators, the approaches obtain results that agree within roughly estimated uncertainties. These methods were also successfully benchmarked against one another, as well as against quasi-exact diagonalization in light nuclei~\cite{Basili:2020PRC,Novario:2021PRL,Yao:2021PRC}.
The difference between NMEs for $0\nu\beta\beta$ decay calculated with different \emph{ab initio} methods but the same input has been found to give a useful approximation to the inaccuracies caused by truncation in many-body methods. 
These studies make it feasible to carry out uncertainty quantification in the {\em ab initio} prediction of the NMEs of experimentally relevant nuclei.  
 
The second-lightest such nucleus, \nuclide[76]{Ge}, is, along with $^{136}$Xe, one of the two most important isotopes for experimental searches, and is now  within the reach of multiple {\em ab initio} methods. The VS-IMSRG was the first \emph{ab initio} approach to calculate the NME for $^{76}$Ge, using the long-range (LR) transition operator associated with standard light-neutrino exchange~\cite{Belley:2021PRL}. 
The resulting NME, 2.14(9), was 25-45\% smaller than those obtained from phenomenological shell-model calculations. However, the contributions of the recently discovered leading-order short-range (SR) contact transition operator~\cite{Cirigliano2018PRL} and higher-order terms were not evaluated. In this work, we now include these contributions. In particular, we report the results from the IM-GCM calculation and present the first comprehensive uncertainty quantification for the  NME in $^{76}$Ge using strong and weak interactions consistently derived within $\chi$EFT.

\begin{figure}[tb]
\centering 
\includegraphics[width=\linewidth]{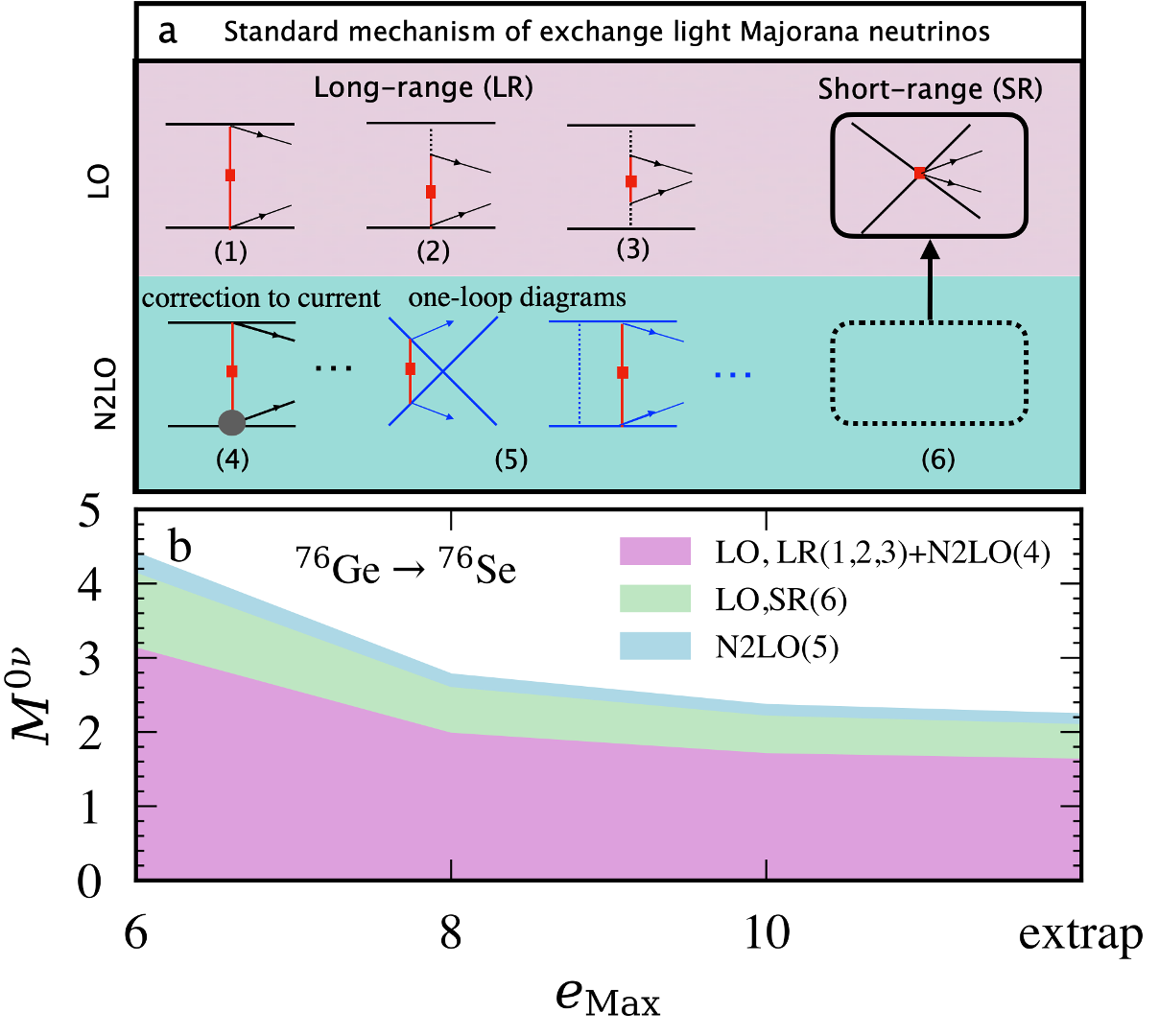} 
\caption{\textbf{Hierarchy of contributions to $0\nu\beta\beta$ decay in chiral EFT, assuming light Majorana neutrino exchange.}  \textbf{a}, Diagrammatic contributions at different orders. \textbf{b}, The convergence of the NMEs for \nuclide[76]{Ge} $\to$ \nuclide[76]{Se} using LO and N2LO transition operators. The red square indicates an insertion of LNV Majorana neutrino mass term $\langle m_{\beta\beta}\rangle$, and the gray circle represents corrections to the single-nucleon current parameterized in terms of form factors. The results are obtained from IM-GCM calculations using the EM1.8/2.0 chiral NN+3N interaction~\cite{Hebeler:2011} as a function of $\eMax$. 
}
\label{fig:IMGCM4NME}
\end{figure}

\paragraph{\textbf{Quantifying the uncertainty in the $0\nu\beta\beta$-decay NME.}}  For the $0\nu\beta\beta$ decay ${}^{76}{\rm Ge}(0^+_1) \to  {}^{76}{\rm Se}(0^+_1) + 2e^-$, the NME, called $M^{0\nu}$, can be written as:
\begin{equation}
 \label{eq:NME}
 M^{0\nu} 
    = \bra{^{76}{\rm Se}(0^+_1)}\hat O^{0\nu}\ket{^{76}{\rm Ge}(0^+_1)}, 
 \end{equation}
where the decay operator $\hat{O}^{0\nu}$ is derived in the standard mechanism of exchange light Majorana neutrinos, depicted in  Fig.~\ref{fig:IMGCM4NME}(a). The wave functions are obtained with the two \emph{ab initio} methods, i.e., IM-GCM and VS-IMSRG. The main challenge in the assessment of theoretical error is the propagation of the uncertainties in the LECs from the chiral interaction through the complicated many-body calculations that ultimately produce the NME. To this end, we use the Sampling/Importance Resampling~\cite{Smith1992} formulation of Bayes' theorem for discrete samples, as was done in Ref.~\cite{Hu:2022} to obtain a theoretical uncertainty on the neutron skin of $^{208}$Pb.

 \begin{table}[bt]
 \centering
 \tabcolsep=6pt
 \caption{\textbf{The recommended value for the total NME of $0\nu\beta\beta$ decay in \nuclide[76]{Ge}, together with the uncertainties from different sources}. } 
 \begin{tabular}{ccccccc}
 \toprule
 $M^{0\nu}$& $\epsilon_{\rm LEC} $& $\epsilon_{\chi \rm EFT}$ & $\epsilon_{\rm MBT}$ & $\epsilon_{\rm OP}$ &$\epsilon_{\rm EM}$  \\
 \hline
 $2.60^{+1.28}_{-1.36}$ & 0.75 & 0.3 &  0.88 &  0.47 &  <0.06  \\
 \bottomrule
 \end{tabular}
 \label{tab:errors}
  \end{table}

Following this procedure, a posterior predictive distribution (PPD) of the NMEs depending on the LECs ($\mathbf{c}$) is given by
\begin{equation}
    \textrm{PPD} = \big\{M^{0\nu}_k(\bold{c}): \bold{c} \sim \mathcal{P} (\mathbf{c}|\textrm{calibration})\big\},
\end{equation}
where $M^{0\nu}_k$ represents the NME from a specific theoretical calculation (i.e. using a particular many-body method and operators truncated at order $k$) and $\mathcal{P} (\bold{c}|\textrm{calibration})$ represents the probability of an LEC sample to yield results for a set of calibration observables that match experimental data.  We label the standard deviation coming from this (non-Gaussian) distribution $\epsilon_{\rm LEC}$ to make comparison with other sources of error easier.
% In this work we use the neutron-proton scattering phase shift in the $^{1}S_0$ partial wave at lab energy of 50 MeV as our calibration observable, since it has recently been discovered to correlate strongly with the NMEs  (see Methods). 
As calibration observables, we use properties of nuclei of mass $A=2-4$ and $A=16$ as done in Ref.~\cite{Jiang:2022} to which we add the neutron-proton scattering phase shift in the $^{1}S_0$ partial wave at lab energy of 50 MeV, since it has recently been discovered to correlate strongly with the NMEs~\cite{Belley_inprep}. The NMEs for the LEC samples are then evaluated using the recently developed Multi-output Multi-fidelity Deep Gaussian Process (MM-DGP) emulator~\cite{Pitcher_inprep} for the VS-IMSRG, which allows us to, within minutes, predict the results of billions of many-body calculations that would otherwise take years to perform in full. 
 
We further assume that our errors are normally distributed and mutually independent, such that the true value of the NME in Eq.~(\ref{eq:NME}) can be written as:  
\begin{equation}
    M^{0\nu} = M^{0\nu}_k + \epsilon_{\chi \rm EFT} + \epsilon_{\rm MBT} + \epsilon_{\rm OP} + \epsilon_{\rm EM},
\end{equation}
where $\epsilon_{\chi \rm EFT} $ represents the error coming from truncation of the nuclear forces, $\epsilon_{\rm MBT}$ the error from the many-body method, $\epsilon_{\rm OP}$ the error due to the truncation of the decay operator and finally, $\epsilon_{\rm EM}$ the error on the emulated results. The values of the NME, together with the errors $\epsilon_{i}$ from different sources, are presented in Table~\ref{tab:errors}.  We detail below how each uncertainty is assessed.

 \begin{figure*}[tb] 
    \includegraphics[width=\linewidth]{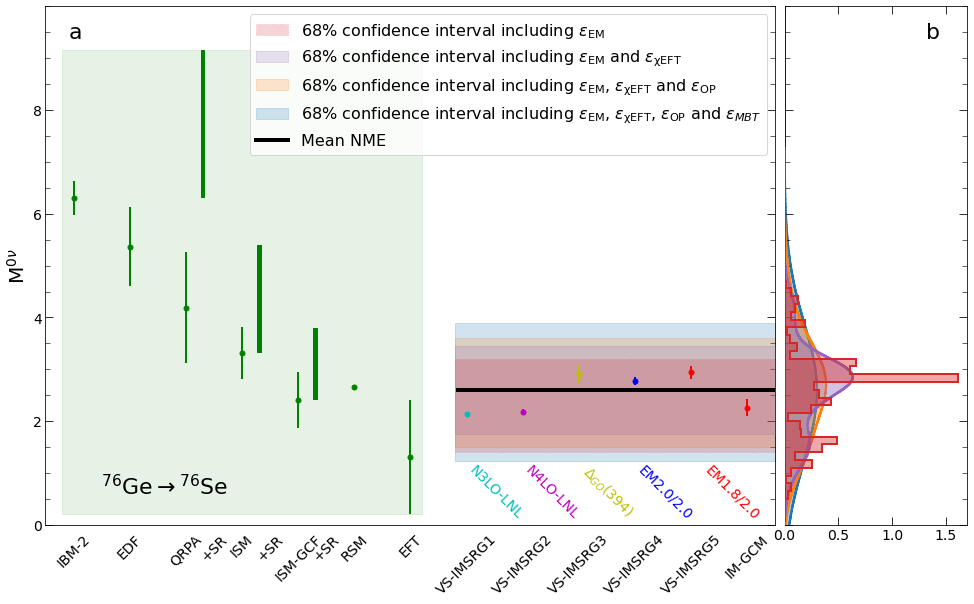}
    \caption{ \textbf{Comparison of $0\nu\beta\beta$-decay NMEs in \nuclide[76]{Ge} from nuclear models and ab initio calculations.}
    \textbf{a}, The NMEs from phenomenological models, including the interacting-boson model (IBM-2)~\cite{Barea:2015,Deppisch:2020ztt}, energy-density-functional (EDF) methods~\cite{Rodriguez:2010,Song:2017}, quasiparticle random-phase approximation (QRPA)~\cite{Mustonen:2013,Hyvarinen:2015,Fang:2018}, interacting shell model (ISM)~\cite{Menendez:2009,Horoi:2016}, ISM with generalized contact formalism (ISM-GCF)~\cite{Weiss:2022}, realistic shell model (RSM)~\cite{Coraggio:2020} and effective field theory (EFT)~\cite{Brase:2021}, are compared to the results of the VS-IMSRG and IM-GCM using different chiral interactions. The error bars of phenomenological nuclear models reflect the discrepancy of calculations from different groups and the bands shows results with the short range contributions included~\cite{Weiss:2022, Jokiniemi:2023}. \textbf{b}, The posterior distribution function of the $0\nu\beta\beta$ NME using the MM-DGP emulator of the VS-IMSRG with 8188 non-implausible samples of chiral interactions from which the confidence intervals are extracted. The final distribution including all errors yields a value of $M^{0\nu\beta\beta}$ = $2.60^{+1.28}_{-1.36}$. The samples are weighted by phase-shifts in the $^{1}S_0$ partial wave and nuclear observables for mass A=2-4,16 as described in the supplemental material~\cite{SupplementalMaterial} . The uncertainties $\epsilon_i$ from EFT, many-body, operator and emulator are then added independently. See text for details.} 
    \label{fig:TotalNME_uncertainty}
\end{figure*}

We employ nuclear interactions derived in a formulation of $\chi {\rm EFT}$ where the $\Delta$-isobars are considered explicitly~\cite{Jiang:2020}. 
In particular, these interactions are given at next-to-next-to-leading order (N2LO) in the chiral expansion, where 17 LECs arise. These interactions are particularly useful for the present study since more diagrammatic contributions are considered at a given order in a $\Delta$-full theory than in a $\Delta$-less one, and the LECs come out more natural. 

To assess $\epsilon_{\chi \rm EFT}$, we need to estimate the contributions coming from neglected higher orders. We do so using a recently developed technique~\cite{Melendez2019}, where, by examining the order-by-order NME convergence, we obtain a 68\% degree of belief interval of 0.27 with the $\Delta$-full interaction, upon which the interactions in this work are based~\cite{Jiang:2020}.
To be conservative, we choose $\epsilon_{\chi \rm EFT}$ to be 0.3, as other parameterizations might have a slower rate of convergence. The results are shown in the supplemental material~\cite{SupplementalMaterial}, compared to the convergence of particular $\Delta$-less nuclear interactions, available at NLO, N2LO, N3LO, and N4LO~\cite{Entem2017, Konstantinos:2020PRC}. As expected, we find more rapid convergence for the $\Delta$-full interaction. Additionally, we observe the change from N3LO to N4LO to be much smaller, indicating that a future analysis with nuclear interactions at N3LO would allow us to greatly reduce that uncertainty.

To estimate $\epsilon_{\rm MBT}$, two \emph{ab initio} methods, i.e., IM-GCM~\cite{Yao:2020PRL} and VS-IMSRG~\cite{Stroberg:2019} are employed. 
These calculations are carried out using the NN+3N chiral interaction, denoted EM1.8/2.0~\cite{Simonis:2017}, where the previous VS-IMSRG value~\cite{Belley:2021PRL} was found to be $M^{0\nu}_{\rm LR}$ = 2.14(9). 
Here we also compute the contribution of the SR contact transition operator and find an overall $\sim$40\% increase in the NME to $M^{0\nu}$ = 2.94(8), with uncertainty coming from both the single-particle basis extrapolation to infinity as well as reference state dependence. 
Similar calculations are carried out with the IM-GCM, yielding a long-range NME of $M_{\rm LR}^{0\nu} =1.67$  and total NME of $M^{0\nu} = 2.25(11)$, where the uncertainties are from the extension of model space with cranking configurations.   In both approaches we find that the NMEs are significantly increased by the SR transition operator, confirming that it contributes at LO.  On the other hand, the results show a modest deviation, where the VS-IMSRG NME is $\sim$30\% larger than that from the IM-GCM. We use the largest discrepancy of 0.88 as an estimate of $\epsilon_{\rm MBT}$, which was shown be be a reasonable  approximation to the inaccuracies caused by truncation in light nuclei~\cite{Yao:2021PRC}. This difference is somewhat larger than what was found in $^{48}$Ca~\cite{Yao:2020PRL, Belley:2021PRL}. This can be understood from the fact that the low-lying states in $^{48}$Ca and $^{48}$Ti are relatively simple and the quadrupole collectivity of $^{48}$Ti is adequately captured in both methods. In contrast, the low-lying states of \nuclide[76]{Ge} and \nuclide[76]{Se} exhibit strong shape-coexistence and collectivity, including significant triaxiality~\cite{Ayangeakaa:2023PRC,Henderson:2019,Rodriguez:2017JPG}.
While these collective degrees of freedom are difficult to capture within the VS-IMSRG and other ab initio methods starting from spherical references~\cite{Hend18E2,Stro22E2}, they are efficiently incorporated within the IM-GCM, as can be seen by comparing excitation spectra and electric multipole transitions with the experimental values~\cite{supp}.
We expect that future systematic improvements to the many-body truncations will improve the agreement between the two methods.

The errors from the $0\nu\beta\beta$ decay transition operators, collectively labeled $\epsilon_{\rm OP}$, can be separated into three sources:  the use of the closure approximation for the intermediate odd-odd nucleus, the determination of the LEC of the SR transition operator, and the truncation of contributions beyond LO in the operator expansion. The potential error stemming from the use of the closure approximation has been assessed with phenomenological nuclear models~\cite{Senkov2013,Wang:2021} to be around 10\% of the LR NME. This finding aligns with the expectation that contributions depending on the excitation energies of intermediate states belong to the N2LO~\cite{Cirigliano:2018PRC}.   Figure~\ref{fig:IMGCM4NME}(a) presents different contribution to the $0\nu\beta\beta$-decay operators at LO and N2LO, noting that there is no contribution at NLO. Figure~\ref{fig:IMGCM4NME}(b) displays the convergence of the NMEs at LO and N2LO with respect to $\eMax$, the number of harmonic oscillator major shells in the basis, in the IM-GCM calculation~\cite{supp}. The value of the LEC for the SR transition operator is determined by fitting the transition amplitude of $nn\to ppe^-e^-$ process following Ref.~\cite{Wirth:2021}. The SR transition contribute is found to be $M^{0\nu}_{\rm SR}=[0.399, 0.526]$, where the about 27\% uncertainty is propagated from the synthetic datum~\cite{Cirigliano:2021PRL}. 
We compute the contribution of the genuine N2LO transition operators, c.f. Fig.~\ref{fig:IMGCM4NME}, which cannot be absorbed into the form factors, while excluding the contributions requiring intermediate states of odd-odd nucleus.  The correction of transition operators at N2LO to the NME shows a weak dependence on the renormalization scale $\mu$, and is found to be 0.079 at $\mu=500$~MeV, consistent with previous findings~\cite{Pastore:2018}. 
This confirms that the N2LO contributions are small, and the power counting works well for the transition operators. 
It also suggests that the common practice of taking LO transition operators supplemented with dipole form factors is a good approximation, once the contact term is properly considered. In short, we take a conservative value $\epsilon_{\rm OP}=0.47$ which includes  0.26 from the use of closure approximation, 0.13 from the uncertainty of the LEC of the SR transition operator, and 0.08 from the truncation on the chiral expansion of transition operators.

Finally, $\epsilon_{\rm EM}$ is given by the MM-DGP emulator as it is based upon Gaussian Processes, which inherently come with a variance for each prediction. 
We obtain the final predictive posterior distribution by sampling the PPD $10^8$ times and adding errors independently sampled from a normal distribution for each $\epsilon$ term.
Figure~\ref{fig:TotalNME_uncertainty} shows the PPDs obtained with each error term discussed above, added separately. 
We find that $M^{0\nu} = 2.60^{+1.28}_{-1.36}$, where the uncertainty represents a 68\% confidence interval. 
We compare the PPD with results obtained from the VS-IMSRG and IM-GCM methods, using the EM1.8/2.0 nuclear interaction~\cite{Hebeler:2011} and VS-IMSRG with four other state-of-the-art chiral NN+3N interactions~\cite{Hebeler:2011,Jiang:2020,Soma:2020}. All these fall within our confidence interval. Our predictions are further compared to NMEs from various phenomenological nuclear models, where the contribution of the contact transition operator is usually not considered due to the challenge in determining the unknown LEC of the SR transition operator in such approaches. With the LECs's value estimated by considering the  charge-independence-breaking coupling of nuclear Hamiltonians, the contribution of the SR operator was quantified with the interacting shell-model (ISM)  and quasiparticle random-phase approximation (QRPA)~\cite{Jokiniemi:2023}. Taking this into account, the discrepancy among different phenomenological models can exceed one order of magnitude, as depicted in Fig.\ref{fig:TotalNME_uncertainty}.

\paragraph{\textbf{Conclusions.}}  
In summary, we have presented the first comprehensive  uncertainty quantification in \emph{ab initio} calculations of NMEs for the $0\nu\beta\beta$ decay of \nuclide[76]{Ge}  using nuclear interactions derived from $\chi {\rm EFT}$ and recently developed many-body emulators based on the standard mechanism of exchanging light Majorana neutrinos with transition operators truncated up to the N2LO.  We have demonstrated that the NME converges rapidly with the chiral expansion, both for the transition operators and for the strong interactions. Considering the uncertainties stemming from different selections of chiral interactions and many-body solvers, our recommended value for the NME stands at $2.60^{+1.28}_{-1.36}$ (68\% C.L.). This, in conjunction with the best  half-life sensitivity of $T_{1/2} > 1.8 \times 10^{26}$ years~\cite{GERDA:Agostini-PRL2020}, phase-space factor of $G_{0\nu}=0.237\times 10^{14}$ years$^{-1}$~\cite{Kotila:2012,Stefanik:2015}, and the axial-vector coupling strength $g_A=1.27$, sets the current upper limit for the effective neutrino mass at $187^{+205}_{-62}$ meV. It is important to note that the current uncertainty encompasses estimated errors from both operators and many-body solvers, presumed to be mutually independent, serving as a reasonable initial framework.  While this uncertainty remains substantial, an effective strategy to mitigate it is now available by considering nuclear interactions that go to higher order in the chiral expansion, reducing many-body truncation errors and improving the likelihood function with a few more relevant observables.  With our NME,  the next-generation tonne-scale Germanium experiment with the ability to discover $0\nu\beta\beta$ decay up to $1.3\times 10^{28}$ years~\cite{LEGEND:Abgrall-2021}  will set the upper limit on effective neutrino mass $\langle m_{\beta\beta}\rangle=22^{+24}_{-7}$ meV, which encompasses almost the entire range of allowed values of inverted neutrino mass hierarchy. This paper complements recent NME calculations in heavy systems~\cite{Belley2023TeXe}, illustrating the power of ab initio methods to potentially deliver quantified uncertainties for all key isotopes of experimental interest. 

\begin{acknowledgments}
\paragraph{Acknowledgments.}
We thank J. de Vries, G. Li, C. F. Jiao, J. Menéndez,  J. Meng, E. Mereghetti, and R. Wirth for fruitful discussions. We thank C. Forssén for providing the non-implausible interactions samples. This work is supported in part by the National Natural Science Foundation of China (Grant Nos. 12375119 and 12141501),  the Guangdong Basic and Applied Basic Research Foundation (2023A1515010936),  the U.S.\ Department of Energy, Office of Science, Office of Nuclear Physics under Awards No. DE-SC0023516, No.\ DE-SC0023175 (NUCLEI SciDAC-5 Collaboration), DE-FG02-97ER41014, DE-FG02-97ER41019, DE-AC02-06CH11357, and DE-SC0015376 (the DBD Topical Theory Collaboration), NSERC, the Arthur B.\ McDonald Canadian Astroparticle Physics Research Institute, the Canadian Institute for Nuclear Physics, the Deutsche Forschungsgemeinschaft (DFG, German Research Foundation) -- Project-ID 279384907 -- SFB 1245, and the European Research Council (ERC) under the European Union’s Horizon 2020 research and innovation programme (Grant Agreement No.\ 101020842). TRIUMF receives funding via a contribution through the National Research Council of Canada. A.M.R. acknowledges the support from NextGenerationEU/PRTR funding. The work of T.R.R. was funded by the Spanish MCIN under contracts PGC2018-094583-B-I00 and PID2021-127890NB-I00. The IM-GCM calculations were carried out using the computing resources provided by the Institute for Cyber-Enabled Research at Michigan State University, the GENCI-TGCC (Contract No.~A0130513012) and CCRT HPC (TOPAZE supercomputer) at Bruyères-le-Châtel, and the U.S.~National Energy Research Scientific Computing Center (NERSC), a DOE Office of Science User Facility supported by the Office of Science of the U.S.~Department of Energy under Contract No.\ DE-AC02-05CH11231.   The VS-IMSRG computations were performed with an allocation of computing resources on Cedar at WestGrid and Compute Canada, and on the Oak Cluster at TRIUMF managed by the University of British Columbia Department of Advanced Research Computing (ARC).

A. B. and J. M. Y. should be considered co-first authors.
\end{acknowledgments}

%\bibliography{abbrev,lib,extra}

\begin{thebibliography}{63}%
\makeatletter
\providecommand \@ifxundefined [1]{%
 \@ifx{#1\undefined}
}%
\providecommand \@ifnum [1]{%
 \ifnum #1\expandafter \@firstoftwo
 \else \expandafter \@secondoftwo
 \fi
}%
\providecommand \@ifx [1]{%
 \ifx #1\expandafter \@firstoftwo
 \else \expandafter \@secondoftwo
 \fi
}%
\providecommand \natexlab [1]{#1}%
\providecommand \enquote  [1]{``#1''}%
\providecommand \bibnamefont  [1]{#1}%
\providecommand \bibfnamefont [1]{#1}%
\providecommand \citenamefont [1]{#1}%
\providecommand \href@noop [0]{\@secondoftwo}%
\providecommand \href [0]{\begingroup \@sanitize@url \@href}%
\providecommand \@href[1]{\@@startlink{#1}\@@href}%
\providecommand \@@href[1]{\endgroup#1\@@endlink}%
\providecommand \@sanitize@url [0]{\catcode `\\12\catcode `\$12\catcode
  `\&12\catcode `\#12\catcode `\^12\catcode `\_12\catcode `\%12\relax}%
\providecommand \@@startlink[1]{}%
\providecommand \@@endlink[0]{}%
\providecommand \url  [0]{\begingroup\@sanitize@url \@url }%
\providecommand \@url [1]{\endgroup\@href {#1}{\urlprefix }}%
\providecommand \urlprefix  [0]{URL }%
\providecommand \Eprint [0]{\href }%
\providecommand \doibase [0]{https://doi.org/}%
\providecommand \selectlanguage [0]{\@gobble}%
\providecommand \bibinfo  [0]{\@secondoftwo}%
\providecommand \bibfield  [0]{\@secondoftwo}%
\providecommand \translation [1]{[#1]}%
\providecommand \BibitemOpen [0]{}%
\providecommand \bibitemStop [0]{}%
\providecommand \bibitemNoStop [0]{.\EOS\space}%
\providecommand \EOS [0]{\spacefactor3000\relax}%
\providecommand \BibitemShut  [1]{\csname bibitem#1\endcsname}%
\let\auto@bib@innerbib\@empty
%</preamble>
\bibitem [{\citenamefont {Fukugita}\ and\ \citenamefont
  {Yanagida}(1986)}]{Fukugita:1986PLB}%
  \BibitemOpen
  \bibfield  {author} {\bibinfo {author} {\bibfnamefont {M.}~\bibnamefont
  {Fukugita}}\ and\ \bibinfo {author} {\bibfnamefont {T.}~\bibnamefont
  {Yanagida}},\ }\bibfield  {title} {\bibinfo {title} {{Baryogenesis Without
  Grand Unification}},\ }\href {https://doi.org/10.1016/0370-2693(86)91126-3}
  {\bibfield  {journal} {\bibinfo  {journal} {Phys. Lett.}\ }\textbf {\bibinfo
  {volume} {B174}},\ \bibinfo {pages} {45} (\bibinfo {year}
  {1986})}\BibitemShut {NoStop}%
%%CITATION = PHLTA,B174,45;%%
\bibitem [{\citenamefont {Furry}(1939)}]{Furry:1939}%
  \BibitemOpen
  \bibfield  {author} {\bibinfo {author} {\bibfnamefont {W.~H.}\ \bibnamefont
  {Furry}},\ }\bibfield  {title} {\bibinfo {title} {{On Transition
  Probabilities in Double Beta-Disintegration}},\ }\href
  {https://doi.org/10.1103/PhysRev.56.1184} {\bibfield  {journal} {\bibinfo
  {journal} {Phys. Rev.}\ }\textbf {\bibinfo {volume} {56}},\ \bibinfo {pages}
  {1184} (\bibinfo {year} {1939})}\BibitemShut {NoStop}%
\bibitem [{\citenamefont {Schechter}\ and\ \citenamefont
  {Valle}(1982)}]{Schechter:1982PRD}%
  \BibitemOpen
  \bibfield  {author} {\bibinfo {author} {\bibfnamefont {J.}~\bibnamefont
  {Schechter}}\ and\ \bibinfo {author} {\bibfnamefont {J.~W.~F.}\ \bibnamefont
  {Valle}},\ }\bibfield  {title} {\bibinfo {title} {{Neutrinoless
  double-$\ensuremath{\beta}$ decay in
  SU(2)\ifmmode\times\else\texttimes\fi{}U(1) theories}},\ }\href
  {https://doi.org/10.1103/PhysRevD.25.2951} {\bibfield  {journal} {\bibinfo
  {journal} {Phys. Rev. D}\ }\textbf {\bibinfo {volume} {25}},\ \bibinfo
  {pages} {2951} (\bibinfo {year} {1982})}\BibitemShut {NoStop}%
\bibitem [{\citenamefont {Agostini}\ and\ \citenamefont {the
  others}(2020)\citenamefont {Agostini} \emph
  {et~al.}}]{GERDA:Agostini-PRL2020}%
  \BibitemOpen
  \bibfield  {author} {\bibinfo {author} {\bibfnamefont {M.}~\bibnamefont
  {Agostini}} \emph {et~al.} (\bibinfo {collaboration} {GERDA Collaboration}),\
  }\bibfield  {title} {\bibinfo {title} {{Final Results of GERDA on the Search
  for Neutrinoless Double-$\ensuremath{\beta}$ Decay}},\ }\href
  {https://doi.org/10.1103/PhysRevLett.125.252502} {\bibfield  {journal}
  {\bibinfo  {journal} {Phys. Rev. Lett.}\ }\textbf {\bibinfo {volume} {125}},\
  \bibinfo {pages} {252502} (\bibinfo {year} {2020})}\BibitemShut {NoStop}%
\bibitem [{\citenamefont {Arnquist}\ and\ \citenamefont {the
  others}(2023)\citenamefont {Arnquist} \emph {et~al.}}]{Majorana:2022}%
  \BibitemOpen
  \bibfield  {author} {\bibinfo {author} {\bibfnamefont {I.~J.}\ \bibnamefont
  {Arnquist}} \emph {et~al.} (\bibinfo {collaboration} {Majorana
  Collaboration}),\ }\bibfield  {title} {\bibinfo {title} {Final result of the
  majorana demonstrator's search for neutrinoless double-$\ensuremath{\beta}$
  decay in $^{76}\mathrm{Ge}$},\ }\href
  {https://doi.org/10.1103/PhysRevLett.130.062501} {\bibfield  {journal}
  {\bibinfo  {journal} {Phys. Rev. Lett.}\ }\textbf {\bibinfo {volume} {130}},\
  \bibinfo {pages} {062501} (\bibinfo {year} {2023})}\BibitemShut {NoStop}%
\bibitem [{\citenamefont {Zhang}\ \emph {et~al.}(2023)\citenamefont {Zhang}
  \emph {et~al.}}]{CDEX:2023}%
  \BibitemOpen
  \bibfield  {author} {\bibinfo {author} {\bibfnamefont {B.~T.}\ \bibnamefont
  {Zhang}} \emph {et~al.} (\bibinfo {collaboration} {CDEX}),\ }\bibfield
  {title} {\bibinfo {title} {{Searching for $^{76}$Ge neutrinoless double beta
  decay with the CDEX-1B experiment}},\ }\href@noop {} {\  (\bibinfo {year}
  {2023})},\ \Eprint {https://arxiv.org/abs/2305.00894} {arXiv:2305.00894
  [nucl-ex]} \BibitemShut {NoStop}%
\bibitem [{\citenamefont {Men\'endez}\ \emph {et~al.}(2009)\citenamefont
  {Men\'endez}, \citenamefont {Poves}, \citenamefont {Caurier},\ and\
  \citenamefont {Nowacki}}]{Menendez:2009}%
  \BibitemOpen
  \bibfield  {author} {\bibinfo {author} {\bibfnamefont {J.}~\bibnamefont
  {Men\'endez}}, \bibinfo {author} {\bibfnamefont {A.}~\bibnamefont {Poves}},
  \bibinfo {author} {\bibfnamefont {E.}~\bibnamefont {Caurier}},\ and\ \bibinfo
  {author} {\bibfnamefont {F.}~\bibnamefont {Nowacki}},\ }\bibfield  {title}
  {\bibinfo {title} {Disassembling the nuclear matrix elements of the
  neutrinoless $\ensuremath{\beta}\ensuremath{\beta}$ decay},\ }\href
  {https://doi.org/10.1016/j.nuclphysa.2008.12.005} {\bibfield  {journal}
  {\bibinfo  {journal} {Nucl. Phys. A}\ }\textbf {\bibinfo {volume} {818}},\
  \bibinfo {pages} {139 } (\bibinfo {year} {2009})}\BibitemShut {NoStop}%
\bibitem [{\citenamefont {Rodr\'{\i}guez}\ and\ \citenamefont
  {Mart\'{\i}nez-Pinedo}(2010)}]{Rodriguez:2010}%
  \BibitemOpen
  \bibfield  {author} {\bibinfo {author} {\bibfnamefont {T.~R.}\ \bibnamefont
  {Rodr\'{\i}guez}}\ and\ \bibinfo {author} {\bibfnamefont {G.}~\bibnamefont
  {Mart\'{\i}nez-Pinedo}},\ }\bibfield  {title} {\bibinfo {title} {Energy
  density functional study of nuclear matrix elements for neutrinoless
  $\ensuremath{\beta}\ensuremath{\beta}$ decay},\ }\href
  {https://doi.org/10.1103/PhysRevLett.105.252503} {\bibfield  {journal}
  {\bibinfo  {journal} {Phys. Rev. Lett.}\ }\textbf {\bibinfo {volume} {105}},\
  \bibinfo {pages} {252503} (\bibinfo {year} {2010})}\BibitemShut {NoStop}%
\bibitem [{\citenamefont {Barea}\ \emph {et~al.}(2015)\citenamefont {Barea},
  \citenamefont {Kotila},\ and\ \citenamefont {Iachello}}]{Barea:2015}%
  \BibitemOpen
  \bibfield  {author} {\bibinfo {author} {\bibfnamefont {J.}~\bibnamefont
  {Barea}}, \bibinfo {author} {\bibfnamefont {J.}~\bibnamefont {Kotila}},\ and\
  \bibinfo {author} {\bibfnamefont {F.}~\bibnamefont {Iachello}},\ }\bibfield
  {title} {\bibinfo {title}
  {$0\ensuremath{\nu}\ensuremath{\beta}\ensuremath{\beta}$ and
  $2\ensuremath{\nu}\ensuremath{\beta}\ensuremath{\beta}$ nuclear matrix
  elements in the interacting boson model with isospin restoration},\ }\href
  {https://doi.org/10.1103/PhysRevC.91.034304} {\bibfield  {journal} {\bibinfo
  {journal} {Phys. Rev. C}\ }\textbf {\bibinfo {volume} {91}},\ \bibinfo
  {pages} {034304} (\bibinfo {year} {2015})}\BibitemShut {NoStop}%
\bibitem [{\citenamefont {Horoi}\ and\ \citenamefont
  {Neacsu}(2016)}]{Horoi:2016}%
  \BibitemOpen
  \bibfield  {author} {\bibinfo {author} {\bibfnamefont {M.}~\bibnamefont
  {Horoi}}\ and\ \bibinfo {author} {\bibfnamefont {A.}~\bibnamefont {Neacsu}},\
  }\bibfield  {title} {\bibinfo {title} {Shell model predictions for
  $^{124}\mathrm{Sn}$ double-$\ensuremath{\beta}$ decay},\ }\href
  {https://doi.org/10.1103/PhysRevC.93.024308} {\bibfield  {journal} {\bibinfo
  {journal} {Phys. Rev. C}\ }\textbf {\bibinfo {volume} {93}},\ \bibinfo
  {pages} {024308} (\bibinfo {year} {2016})}\BibitemShut {NoStop}%
\bibitem [{\citenamefont {Song}\ \emph {et~al.}(2017)\citenamefont {Song},
  \citenamefont {Yao}, \citenamefont {Ring},\ and\ \citenamefont
  {Meng}}]{Song:2017}%
  \BibitemOpen
  \bibfield  {author} {\bibinfo {author} {\bibfnamefont {L.~S.}\ \bibnamefont
  {Song}}, \bibinfo {author} {\bibfnamefont {J.~M.}\ \bibnamefont {Yao}},
  \bibinfo {author} {\bibfnamefont {P.}~\bibnamefont {Ring}},\ and\ \bibinfo
  {author} {\bibfnamefont {J.}~\bibnamefont {Meng}},\ }\bibfield  {title}
  {\bibinfo {title} {Nuclear matrix element of neutrinoless
  double-$\ensuremath{\beta}$ decay: Relativity and short-range correlations},\
  }\href {https://doi.org/10.1103/PhysRevC.95.024305} {\bibfield  {journal}
  {\bibinfo  {journal} {Phys. Rev. C}\ }\textbf {\bibinfo {volume} {95}},\
  \bibinfo {pages} {024305} (\bibinfo {year} {2017})}\BibitemShut {NoStop}%
\bibitem [{\citenamefont {Fang}\ \emph {et~al.}(2018)\citenamefont {Fang},
  \citenamefont {Faessler},\ and\ \citenamefont {\ifmmode~\check{S}\else
  \v{S}\fi{}imkovic}}]{Fang:2018}%
  \BibitemOpen
  \bibfield  {author} {\bibinfo {author} {\bibfnamefont {D.-L.}\ \bibnamefont
  {Fang}}, \bibinfo {author} {\bibfnamefont {A.}~\bibnamefont {Faessler}},\
  and\ \bibinfo {author} {\bibfnamefont {F.}~\bibnamefont
  {\ifmmode~\check{S}\else \v{S}\fi{}imkovic}},\ }\bibfield  {title} {\bibinfo
  {title} {$0\ensuremath{\nu}\ensuremath{\beta}\ensuremath{\beta}$-decay
  nuclear matrix element for light and heavy neutrino mass mechanisms from
  deformed quasiparticle random-phase approximation calculations for
  $^{76}\mathrm{Ge}, ^{82}\mathrm{Se}, ^{130}\mathrm{Te}, ^{136}\mathrm{Xe}$,
  and $^{150}\mathrm{Nd}$ with isospin restoration},\ }\href
  {https://doi.org/10.1103/PhysRevC.97.045503} {\bibfield  {journal} {\bibinfo
  {journal} {Phys. Rev. C}\ }\textbf {\bibinfo {volume} {97}},\ \bibinfo
  {pages} {045503} (\bibinfo {year} {2018})}\BibitemShut {NoStop}%
\bibitem [{\citenamefont {Coraggio}\ \emph {et~al.}(2020)\citenamefont
  {Coraggio}, \citenamefont {Gargano}, \citenamefont {Itaco}, \citenamefont
  {Mancino},\ and\ \citenamefont {Nowacki}}]{Coraggio:2020}%
  \BibitemOpen
  \bibfield  {author} {\bibinfo {author} {\bibfnamefont {L.}~\bibnamefont
  {Coraggio}}, \bibinfo {author} {\bibfnamefont {A.}~\bibnamefont {Gargano}},
  \bibinfo {author} {\bibfnamefont {N.}~\bibnamefont {Itaco}}, \bibinfo
  {author} {\bibfnamefont {R.}~\bibnamefont {Mancino}},\ and\ \bibinfo {author}
  {\bibfnamefont {F.}~\bibnamefont {Nowacki}},\ }\bibfield  {title} {\bibinfo
  {title} {Calculation of the neutrinoless double-$\ensuremath{\beta}$ decay
  matrix element within the realistic shell model},\ }\href
  {https://doi.org/10.1103/PhysRevC.101.044315} {\bibfield  {journal} {\bibinfo
   {journal} {Phys. Rev. C}\ }\textbf {\bibinfo {volume} {101}},\ \bibinfo
  {pages} {044315} (\bibinfo {year} {2020})}\BibitemShut {NoStop}%
\bibitem [{\citenamefont {Engel}(2015)}]{Engel:2015}%
  \BibitemOpen
  \bibfield  {author} {\bibinfo {author} {\bibfnamefont {J.}~\bibnamefont
  {Engel}},\ }\bibfield  {title} {\bibinfo {title} {Uncertainties in nuclear
  matrix elements for neutrinoless double-beta decay},\ }\href
  {https://doi.org/10.1088/0954-3899/42/3/034017} {\bibfield  {journal}
  {\bibinfo  {journal} {Journal of Physics G: Nuclear and Particle Physics}\
  }\textbf {\bibinfo {volume} {42}},\ \bibinfo {pages} {034017} (\bibinfo
  {year} {2015})}\BibitemShut {NoStop}%
\bibitem [{\citenamefont {Engel}\ and\ \citenamefont
  {Men{\'{e}}ndez}(2017)}]{Engel:2017}%
  \BibitemOpen
  \bibfield  {author} {\bibinfo {author} {\bibfnamefont {J.}~\bibnamefont
  {Engel}}\ and\ \bibinfo {author} {\bibfnamefont {J.}~\bibnamefont
  {Men{\'{e}}ndez}},\ }\bibfield  {title} {\bibinfo {title} {Status and future
  of nuclear matrix elements for neutrinoless double-beta decay: a review},\
  }\href {https://doi.org/10.1088/1361-6633/aa5bc5} {\bibfield  {journal}
  {\bibinfo  {journal} {Rep. Prog. Phys.}\ }\textbf {\bibinfo {volume} {80}},\
  \bibinfo {pages} {046301} (\bibinfo {year} {2017})}\BibitemShut {NoStop}%
\bibitem [{\citenamefont {Yao}\ \emph {et~al.}(2022)\citenamefont {Yao},
  \citenamefont {Meng}, \citenamefont {Niu},\ and\ \citenamefont
  {Ring}}]{Yao:2022PPNP}%
  \BibitemOpen
  \bibfield  {author} {\bibinfo {author} {\bibfnamefont {J.~M.}\ \bibnamefont
  {Yao}}, \bibinfo {author} {\bibfnamefont {J.}~\bibnamefont {Meng}}, \bibinfo
  {author} {\bibfnamefont {Y.~F.}\ \bibnamefont {Niu}},\ and\ \bibinfo {author}
  {\bibfnamefont {P.}~\bibnamefont {Ring}},\ }\bibfield  {title} {\bibinfo
  {title} {{Beyond-mean-field approaches for nuclear neutrinoless double beta
  decay in the standard mechanism}},\ }\href
  {https://doi.org/10.1016/j.ppnp.2022.103965} {\bibfield  {journal} {\bibinfo
  {journal} {Prog. Part. Nucl. Phys.}\ }\textbf {\bibinfo {volume} {126}},\
  \bibinfo {pages} {103965} (\bibinfo {year} {2022})}\BibitemShut {NoStop}%
\bibitem [{\citenamefont {Agostini}\ \emph {et~al.}(2023)\citenamefont
  {Agostini}, \citenamefont {Benato}, \citenamefont {Detwiler}, \citenamefont
  {Men\'endez},\ and\ \citenamefont {Vissani}}]{Agostini:2022Review}%
  \BibitemOpen
  \bibfield  {author} {\bibinfo {author} {\bibfnamefont {M.}~\bibnamefont
  {Agostini}}, \bibinfo {author} {\bibfnamefont {G.}~\bibnamefont {Benato}},
  \bibinfo {author} {\bibfnamefont {J.~A.}\ \bibnamefont {Detwiler}}, \bibinfo
  {author} {\bibfnamefont {J.}~\bibnamefont {Men\'endez}},\ and\ \bibinfo
  {author} {\bibfnamefont {F.}~\bibnamefont {Vissani}},\ }\bibfield  {title}
  {\bibinfo {title} {Toward the discovery of matter creation with neutrinoless
  $\ensuremath{\beta}\ensuremath{\beta}$ decay},\ }\href
  {https://doi.org/10.1103/RevModPhys.95.025002} {\bibfield  {journal}
  {\bibinfo  {journal} {Rev. Mod. Phys.}\ }\textbf {\bibinfo {volume} {95}},\
  \bibinfo {pages} {025002} (\bibinfo {year} {2023})}\BibitemShut {NoStop}%
\bibitem [{\citenamefont {Epelbaum}\ \emph {et~al.}(2009)\citenamefont
  {Epelbaum}, \citenamefont {Hammer},\ and\ \citenamefont
  {Meissner}}]{Epelbaum:2009RMP}%
  \BibitemOpen
  \bibfield  {author} {\bibinfo {author} {\bibfnamefont {E.}~\bibnamefont
  {Epelbaum}}, \bibinfo {author} {\bibfnamefont {H.-W.}\ \bibnamefont
  {Hammer}},\ and\ \bibinfo {author} {\bibfnamefont {U.-G.}\ \bibnamefont
  {Meissner}},\ }\bibfield  {title} {\bibinfo {title} {Modern theory of nuclear
  forces},\ }\href {https://doi.org/10.1103/RevModPhys.81.1773} {\bibfield
  {journal} {\bibinfo  {journal} {Rev. Mod. Phys.}\ }\textbf {\bibinfo {volume}
  {81}},\ \bibinfo {pages} {1773} (\bibinfo {year} {2009})}\BibitemShut
  {NoStop}%
\bibitem [{\citenamefont {Machleidt}\ and\ \citenamefont
  {Entem}(2011)}]{Machleidt:2011PR}%
  \BibitemOpen
  \bibfield  {author} {\bibinfo {author} {\bibfnamefont {R.}~\bibnamefont
  {Machleidt}}\ and\ \bibinfo {author} {\bibfnamefont {D.~R.}\ \bibnamefont
  {Entem}},\ }\bibfield  {title} {\bibinfo {title} {{Chiral effective field
  theory and nuclear forces}},\ }\href
  {https://doi.org/10.1016/j.physrep.2011.02.001} {\bibfield  {journal}
  {\bibinfo  {journal} {Phys. Rept.}\ }\textbf {\bibinfo {volume} {503}},\
  \bibinfo {pages} {1} (\bibinfo {year} {2011})}\BibitemShut {NoStop}%
\bibitem [{\citenamefont {Bogner}\ \emph {et~al.}(2010)\citenamefont {Bogner},
  \citenamefont {Furnstahl},\ and\ \citenamefont {Schwenk}}]{Bogner:2010}%
  \BibitemOpen
  \bibfield  {author} {\bibinfo {author} {\bibfnamefont {S.}~\bibnamefont
  {Bogner}}, \bibinfo {author} {\bibfnamefont {R.}~\bibnamefont {Furnstahl}},\
  and\ \bibinfo {author} {\bibfnamefont {A.}~\bibnamefont {Schwenk}},\
  }\bibfield  {title} {\bibinfo {title} {From low-momentum interactions to
  nuclear structure},\ }\href {https://doi.org/10.1016/j.ppnp.2010.03.001}
  {\bibfield  {journal} {\bibinfo  {journal} {Prog. Part. Nucl. Phys.}\
  }\textbf {\bibinfo {volume} {65}},\ \bibinfo {pages} {94 } (\bibinfo {year}
  {2010})}\BibitemShut {NoStop}%
\bibitem [{\citenamefont {Stroberg}\ \emph {et~al.}(2021)\citenamefont
  {Stroberg}, \citenamefont {Holt}, \citenamefont {Schwenk},\ and\
  \citenamefont {Simonis}}]{Stroberg:2021PRL}%
  \BibitemOpen
  \bibfield  {author} {\bibinfo {author} {\bibfnamefont {S.~R.}\ \bibnamefont
  {Stroberg}}, \bibinfo {author} {\bibfnamefont {J.~D.}\ \bibnamefont {Holt}},
  \bibinfo {author} {\bibfnamefont {A.}~\bibnamefont {Schwenk}},\ and\ \bibinfo
  {author} {\bibfnamefont {J.}~\bibnamefont {Simonis}},\ }\bibfield  {title}
  {\bibinfo {title} {Ab initio limits of atomic nuclei},\ }\href
  {https://doi.org/10.1103/PhysRevLett.126.022501} {\bibfield  {journal}
  {\bibinfo  {journal} {Phys. Rev. Lett.}\ }\textbf {\bibinfo {volume} {126}},\
  \bibinfo {pages} {022501} (\bibinfo {year} {2021})}\BibitemShut {NoStop}%
\bibitem [{\citenamefont {Hu}\ \emph {et~al.}(2022)\citenamefont {Hu} \emph
  {et~al.}}]{Hu:2022}%
  \BibitemOpen
  \bibfield  {author} {\bibinfo {author} {\bibfnamefont {B.}~\bibnamefont {Hu}}
  \emph {et~al.},\ }\bibfield  {title} {\bibinfo {title} {{Ab initio
  predictions link the neutron skin of $^{208}$Pb to nuclear forces}},\ }\href
  {https://doi.org/10.1038/s41567-022-01715-8} {\bibfield  {journal} {\bibinfo
  {journal} {Nature Phys.}\ }\textbf {\bibinfo {volume} {18}},\ \bibinfo
  {pages} {1196} (\bibinfo {year} {2022})}\BibitemShut {NoStop}%
\bibitem [{\citenamefont {Yao}\ \emph {et~al.}(2020)\citenamefont {Yao},
  \citenamefont {Bally}, \citenamefont {Engel}, \citenamefont {Wirth},
  \citenamefont {Rodr\'{\i}guez},\ and\ \citenamefont {Hergert}}]{Yao:2020PRL}%
  \BibitemOpen
  \bibfield  {author} {\bibinfo {author} {\bibfnamefont {J.~M.}\ \bibnamefont
  {Yao}}, \bibinfo {author} {\bibfnamefont {B.}~\bibnamefont {Bally}}, \bibinfo
  {author} {\bibfnamefont {J.}~\bibnamefont {Engel}}, \bibinfo {author}
  {\bibfnamefont {R.}~\bibnamefont {Wirth}}, \bibinfo {author} {\bibfnamefont
  {T.~R.}\ \bibnamefont {Rodr\'{\i}guez}},\ and\ \bibinfo {author}
  {\bibfnamefont {H.}~\bibnamefont {Hergert}},\ }\bibfield  {title} {\bibinfo
  {title} {Ab initio treatment of collective correlations and the neutrinoless
  double beta decay of $^{48}\mathrm{Ca}$},\ }\href
  {https://doi.org/10.1103/PhysRevLett.124.232501} {\bibfield  {journal}
  {\bibinfo  {journal} {Phys. Rev. Lett.}\ }\textbf {\bibinfo {volume} {124}},\
  \bibinfo {pages} {232501} (\bibinfo {year} {2020})}\BibitemShut {NoStop}%
\bibitem [{\citenamefont {Belley}\ \emph {et~al.}(2021)\citenamefont {Belley},
  \citenamefont {Payne}, \citenamefont {Stroberg}, \citenamefont {Miyagi},\
  and\ \citenamefont {Holt}}]{Belley:2021PRL}%
  \BibitemOpen
  \bibfield  {author} {\bibinfo {author} {\bibfnamefont {A.}~\bibnamefont
  {Belley}}, \bibinfo {author} {\bibfnamefont {C.~G.}\ \bibnamefont {Payne}},
  \bibinfo {author} {\bibfnamefont {S.~R.}\ \bibnamefont {Stroberg}}, \bibinfo
  {author} {\bibfnamefont {T.}~\bibnamefont {Miyagi}},\ and\ \bibinfo {author}
  {\bibfnamefont {J.~D.}\ \bibnamefont {Holt}},\ }\bibfield  {title} {\bibinfo
  {title} {Ab initio neutrinoless double-beta decay matrix elements for
  $^{48}\mathrm{Ca}$, $^{76}\mathrm{Ge}$, and $^{82}\mathrm{Se}$},\ }\href
  {https://doi.org/10.1103/PhysRevLett.126.042502} {\bibfield  {journal}
  {\bibinfo  {journal} {Phys. Rev. Lett.}\ }\textbf {\bibinfo {volume} {126}},\
  \bibinfo {pages} {042502} (\bibinfo {year} {2021})}\BibitemShut {NoStop}%
\bibitem [{\citenamefont {Novario}\ \emph {et~al.}(2021)\citenamefont
  {Novario}, \citenamefont {Gysbers}, \citenamefont {Engel}, \citenamefont
  {Hagen}, \citenamefont {Jansen}, \citenamefont {Morris}, \citenamefont
  {Navr\'atil}, \citenamefont {Papenbrock},\ and\ \citenamefont
  {Quaglioni}}]{Novario:2021PRL}%
  \BibitemOpen
  \bibfield  {author} {\bibinfo {author} {\bibfnamefont {S.}~\bibnamefont
  {Novario}}, \bibinfo {author} {\bibfnamefont {P.}~\bibnamefont {Gysbers}},
  \bibinfo {author} {\bibfnamefont {J.}~\bibnamefont {Engel}}, \bibinfo
  {author} {\bibfnamefont {G.}~\bibnamefont {Hagen}}, \bibinfo {author}
  {\bibfnamefont {G.~R.}\ \bibnamefont {Jansen}}, \bibinfo {author}
  {\bibfnamefont {T.~D.}\ \bibnamefont {Morris}}, \bibinfo {author}
  {\bibfnamefont {P.}~\bibnamefont {Navr\'atil}}, \bibinfo {author}
  {\bibfnamefont {T.}~\bibnamefont {Papenbrock}},\ and\ \bibinfo {author}
  {\bibfnamefont {S.}~\bibnamefont {Quaglioni}},\ }\bibfield  {title} {\bibinfo
  {title} {Coupled-cluster calculations of neutrinoless
  double-$\ensuremath{\beta}$ decay in $^{48}\mathrm{Ca}$},\ }\href
  {https://doi.org/10.1103/PhysRevLett.126.182502} {\bibfield  {journal}
  {\bibinfo  {journal} {Phys. Rev. Lett.}\ }\textbf {\bibinfo {volume} {126}},\
  \bibinfo {pages} {182502} (\bibinfo {year} {2021})}\BibitemShut {NoStop}%
\bibitem [{\citenamefont {Basili}\ \emph {et~al.}(2020)\citenamefont {Basili},
  \citenamefont {Yao}, \citenamefont {Engel}, \citenamefont {Hergert},
  \citenamefont {Lockner}, \citenamefont {Maris},\ and\ \citenamefont
  {Vary}}]{Basili:2020PRC}%
  \BibitemOpen
  \bibfield  {author} {\bibinfo {author} {\bibfnamefont {R.~A.~M.}\
  \bibnamefont {Basili}}, \bibinfo {author} {\bibfnamefont {J.~M.}\
  \bibnamefont {Yao}}, \bibinfo {author} {\bibfnamefont {J.}~\bibnamefont
  {Engel}}, \bibinfo {author} {\bibfnamefont {H.}~\bibnamefont {Hergert}},
  \bibinfo {author} {\bibfnamefont {M.}~\bibnamefont {Lockner}}, \bibinfo
  {author} {\bibfnamefont {P.}~\bibnamefont {Maris}},\ and\ \bibinfo {author}
  {\bibfnamefont {J.~P.}\ \bibnamefont {Vary}},\ }\bibfield  {title} {\bibinfo
  {title} {Benchmark neutrinoless double-$\ensuremath{\beta}$ decay matrix
  elements in a light nucleus},\ }\href
  {https://doi.org/10.1103/PhysRevC.102.014302} {\bibfield  {journal} {\bibinfo
   {journal} {Phys. Rev. C}\ }\textbf {\bibinfo {volume} {102}},\ \bibinfo
  {pages} {014302} (\bibinfo {year} {2020})}\BibitemShut {NoStop}%
\bibitem [{\citenamefont {Yao}\ \emph {et~al.}(2021)\citenamefont {Yao},
  \citenamefont {Belley}, \citenamefont {Wirth}, \citenamefont {Miyagi},
  \citenamefont {Payne}, \citenamefont {Stroberg}, \citenamefont {Hergert},\
  and\ \citenamefont {Holt}}]{Yao:2021PRC}%
  \BibitemOpen
  \bibfield  {author} {\bibinfo {author} {\bibfnamefont {J.~M.}\ \bibnamefont
  {Yao}}, \bibinfo {author} {\bibfnamefont {A.}~\bibnamefont {Belley}},
  \bibinfo {author} {\bibfnamefont {R.}~\bibnamefont {Wirth}}, \bibinfo
  {author} {\bibfnamefont {T.}~\bibnamefont {Miyagi}}, \bibinfo {author}
  {\bibfnamefont {C.~G.}\ \bibnamefont {Payne}}, \bibinfo {author}
  {\bibfnamefont {S.~R.}\ \bibnamefont {Stroberg}}, \bibinfo {author}
  {\bibfnamefont {H.}~\bibnamefont {Hergert}},\ and\ \bibinfo {author}
  {\bibfnamefont {J.~D.}\ \bibnamefont {Holt}},\ }\bibfield  {title} {\bibinfo
  {title} {Ab initio benchmarks of neutrinoless double-$\ensuremath{\beta}$
  decay in light nuclei with a chiral hamiltonian},\ }\href
  {https://doi.org/10.1103/PhysRevC.103.014315} {\bibfield  {journal} {\bibinfo
   {journal} {Phys. Rev. C}\ }\textbf {\bibinfo {volume} {103}},\ \bibinfo
  {pages} {014315} (\bibinfo {year} {2021})}\BibitemShut {NoStop}%
\bibitem [{\citenamefont {Cirigliano}\ \emph
  {et~al.}(2018{\natexlab{a}})\citenamefont {Cirigliano}, \citenamefont
  {Dekens}, \citenamefont {de~Vries}, \citenamefont {Graesser}, \citenamefont
  {Mereghetti}, \citenamefont {Pastore},\ and\ \citenamefont {van
  Kolck}}]{Cirigliano2018PRL}%
  \BibitemOpen
  \bibfield  {author} {\bibinfo {author} {\bibfnamefont {V.}~\bibnamefont
  {Cirigliano}}, \bibinfo {author} {\bibfnamefont {W.}~\bibnamefont {Dekens}},
  \bibinfo {author} {\bibfnamefont {J.}~\bibnamefont {de~Vries}}, \bibinfo
  {author} {\bibfnamefont {M.~L.}\ \bibnamefont {Graesser}}, \bibinfo {author}
  {\bibfnamefont {E.}~\bibnamefont {Mereghetti}}, \bibinfo {author}
  {\bibfnamefont {S.}~\bibnamefont {Pastore}},\ and\ \bibinfo {author}
  {\bibfnamefont {U.}~\bibnamefont {van Kolck}},\ }\bibfield  {title} {\bibinfo
  {title} {New leading contribution to neutrinoless double-$\ensuremath{\beta}$
  decay},\ }\href {https://doi.org/10.1103/PhysRevLett.120.202001} {\bibfield
  {journal} {\bibinfo  {journal} {Phys. Rev. Lett.}\ }\textbf {\bibinfo
  {volume} {120}},\ \bibinfo {pages} {202001} (\bibinfo {year}
  {2018}{\natexlab{a}})}\BibitemShut {NoStop}%
\bibitem [{\citenamefont {Smith}\ and\ \citenamefont
  {Gelfand}(1992)}]{Smith1992}%
  \BibitemOpen
  \bibfield  {author} {\bibinfo {author} {\bibfnamefont {A.~F.~M.}\
  \bibnamefont {Smith}}\ and\ \bibinfo {author} {\bibfnamefont {A.~E.}\
  \bibnamefont {Gelfand}},\ }\bibfield  {title} {\bibinfo {title} {{Bayesian
  Statistics without Tears: A Sampling-Resampling Perspective}},\ }\href@noop
  {} {\bibfield  {journal} {\bibinfo  {journal} {Source: The American
  Statistician}\ }\textbf {\bibinfo {volume} {46}},\ \bibinfo {pages} {84}
  (\bibinfo {year} {1992})}\BibitemShut {NoStop}%
\bibitem [{\citenamefont {Jiang}\ \emph {et~al.}(2022)\citenamefont {Jiang},
  \citenamefont {Forss\'en}, \citenamefont {Dj\"arv},\ and\ \citenamefont
  {Hagen}}]{Jiang:2022}%
  \BibitemOpen
  \bibfield  {author} {\bibinfo {author} {\bibfnamefont {W.~G.}\ \bibnamefont
  {Jiang}}, \bibinfo {author} {\bibfnamefont {C.}~\bibnamefont {Forss\'en}},
  \bibinfo {author} {\bibfnamefont {T.}~\bibnamefont {Dj\"arv}},\ and\ \bibinfo
  {author} {\bibfnamefont {G.}~\bibnamefont {Hagen}},\ }\bibfield  {title}
  {\bibinfo {title} {{Emulating ab initio computations of infinite nucleonic
  matter}},\ }\href@noop {} {\  (\bibinfo {year} {2022})},\ \Eprint
  {https://arxiv.org/abs/2212.13216} {arXiv:2212.13216 [nucl-th]} \BibitemShut
  {NoStop}%
\bibitem [{\citenamefont {Belley}\ \emph {et~al.}()\citenamefont {Belley} \emph
  {et~al.}}]{Belley_inprep}%
  \BibitemOpen
  \bibfield  {author} {\bibinfo {author} {\bibfnamefont {A.}~\bibnamefont
  {Belley}} \emph {et~al.},\ }\href@noop {} {}\bibinfo {note} {In
  preparation}\BibitemShut {NoStop}%
\bibitem [{\citenamefont {Pitcher}\ \emph {et~al.}()\citenamefont {Pitcher}
  \emph {et~al.}}]{Pitcher_inprep}%
  \BibitemOpen
  \bibfield  {author} {\bibinfo {author} {\bibfnamefont {J.}~\bibnamefont
  {Pitcher}} \emph {et~al.},\ }\href@noop {} {}\bibinfo {note} {In
  preparation}\BibitemShut {NoStop}%
\bibitem [{\citenamefont {Hebeler}\ \emph {et~al.}(2011)\citenamefont
  {Hebeler}, \citenamefont {Bogner}, \citenamefont {Furnstahl}, \citenamefont
  {Nogga},\ and\ \citenamefont {Schwenk}}]{Hebeler:2011}%
  \BibitemOpen
  \bibfield  {author} {\bibinfo {author} {\bibfnamefont {K.}~\bibnamefont
  {Hebeler}}, \bibinfo {author} {\bibfnamefont {S.~K.}\ \bibnamefont {Bogner}},
  \bibinfo {author} {\bibfnamefont {R.~J.}\ \bibnamefont {Furnstahl}}, \bibinfo
  {author} {\bibfnamefont {A.}~\bibnamefont {Nogga}},\ and\ \bibinfo {author}
  {\bibfnamefont {A.}~\bibnamefont {Schwenk}},\ }\bibfield  {title} {\bibinfo
  {title} {Improved nuclear matter calculations from chiral low-momentum
  interactions},\ }\href {https://doi.org/10.1103/PhysRevC.83.031301}
  {\bibfield  {journal} {\bibinfo  {journal} {Phys. Rev. C}\ }\textbf {\bibinfo
  {volume} {83}},\ \bibinfo {pages} {031301} (\bibinfo {year}
  {2011})}\BibitemShut {NoStop}%
\bibitem [{\citenamefont {Deppisch}\ \emph {et~al.}(2020)\citenamefont
  {Deppisch}, \citenamefont {Graf}, \citenamefont {Iachello},\ and\
  \citenamefont {Kotila}}]{Deppisch:2020ztt}%
  \BibitemOpen
  \bibfield  {author} {\bibinfo {author} {\bibfnamefont {F.~F.}\ \bibnamefont
  {Deppisch}}, \bibinfo {author} {\bibfnamefont {L.}~\bibnamefont {Graf}},
  \bibinfo {author} {\bibfnamefont {F.}~\bibnamefont {Iachello}},\ and\
  \bibinfo {author} {\bibfnamefont {J.}~\bibnamefont {Kotila}},\ }\bibfield
  {title} {\bibinfo {title} {Analysis of light neutrino exchange and
  short-range mechanisms in
  $0\ensuremath{\nu}\ensuremath{\beta}\ensuremath{\beta}$ decay},\ }\href
  {https://doi.org/10.1103/PhysRevD.102.095016} {\bibfield  {journal} {\bibinfo
   {journal} {Phys. Rev. D}\ }\textbf {\bibinfo {volume} {102}},\ \bibinfo
  {pages} {095016} (\bibinfo {year} {2020})}\BibitemShut {NoStop}%
\bibitem [{\citenamefont {Mustonen}\ and\ \citenamefont
  {Engel}(2013)}]{Mustonen:2013}%
  \BibitemOpen
  \bibfield  {author} {\bibinfo {author} {\bibfnamefont {M.~T.}\ \bibnamefont
  {Mustonen}}\ and\ \bibinfo {author} {\bibfnamefont {J.}~\bibnamefont
  {Engel}},\ }\bibfield  {title} {\bibinfo {title} {Large-scale calculations of
  the double-$\beta$ decay of $^{76}\mathrm{Ge}$,$^{130}\mathrm{Te}$,
  $^{136}\mathrm{Xe}$, and $^{150}\mathrm{Nd}$ in the deformed self-consistent
  skyrme quasiparticle random-phase approximation},\ }\href
  {https://doi.org/10.1103/PhysRevC.87.064302} {\bibfield  {journal} {\bibinfo
  {journal} {Phys. Rev. C}\ }\textbf {\bibinfo {volume} {87}},\ \bibinfo
  {pages} {064302} (\bibinfo {year} {2013})}\BibitemShut {NoStop}%
\bibitem [{\citenamefont {Hyv\"arinen}\ and\ \citenamefont
  {Suhonen}(2015)}]{Hyvarinen:2015}%
  \BibitemOpen
  \bibfield  {author} {\bibinfo {author} {\bibfnamefont {J.}~\bibnamefont
  {Hyv\"arinen}}\ and\ \bibinfo {author} {\bibfnamefont {J.}~\bibnamefont
  {Suhonen}},\ }\bibfield  {title} {\bibinfo {title} {Nuclear matrix elements
  for $0\ensuremath{\nu}\ensuremath{\beta}\ensuremath{\beta}$ decays with light
  or heavy majorana-neutrino exchange},\ }\href
  {https://doi.org/10.1103/PhysRevC.91.024613} {\bibfield  {journal} {\bibinfo
  {journal} {Phys. Rev. C}\ }\textbf {\bibinfo {volume} {91}},\ \bibinfo
  {pages} {024613} (\bibinfo {year} {2015})}\BibitemShut {NoStop}%
\bibitem [{\citenamefont {Weiss}\ \emph {et~al.}(2022)\citenamefont {Weiss},
  \citenamefont {Soriano}, \citenamefont {Lovato}, \citenamefont {Menendez},\
  and\ \citenamefont {Wiringa}}]{Weiss:2022}%
  \BibitemOpen
  \bibfield  {author} {\bibinfo {author} {\bibfnamefont {R.}~\bibnamefont
  {Weiss}}, \bibinfo {author} {\bibfnamefont {P.}~\bibnamefont {Soriano}},
  \bibinfo {author} {\bibfnamefont {A.}~\bibnamefont {Lovato}}, \bibinfo
  {author} {\bibfnamefont {J.}~\bibnamefont {Menendez}},\ and\ \bibinfo
  {author} {\bibfnamefont {R.~B.}\ \bibnamefont {Wiringa}},\ }\bibfield
  {title} {\bibinfo {title} {Neutrinoless double-$\ensuremath{\beta}$ decay:
  Combining quantum monte carlo and the nuclear shell model with the
  generalized contact formalism},\ }\href@noop {} {\bibfield  {journal}
  {\bibinfo  {journal} {Phys. Rev. C}\ }\textbf {\bibinfo {volume} {106}},\
  \bibinfo {pages} {065501} (\bibinfo {year} {2022})}\BibitemShut {NoStop}%
\bibitem [{\citenamefont {Brase}\ \emph {et~al.}(2022)\citenamefont {Brase},
  \citenamefont {Men\'endez}, \citenamefont {Coello~P\'erez},\ and\
  \citenamefont {Schwenk}}]{Brase:2021}%
  \BibitemOpen
  \bibfield  {author} {\bibinfo {author} {\bibfnamefont {C.}~\bibnamefont
  {Brase}}, \bibinfo {author} {\bibfnamefont {J.}~\bibnamefont {Men\'endez}},
  \bibinfo {author} {\bibfnamefont {E.~A.}\ \bibnamefont {Coello~P\'erez}},\
  and\ \bibinfo {author} {\bibfnamefont {A.}~\bibnamefont {Schwenk}},\
  }\bibfield  {title} {\bibinfo {title} {Neutrinoless
  double-$\ensuremath{\beta}$ decay from an effective field theory for heavy
  nuclei},\ }\href {https://doi.org/10.1103/PhysRevC.106.034309} {\bibfield
  {journal} {\bibinfo  {journal} {Phys. Rev. C}\ }\textbf {\bibinfo {volume}
  {106}},\ \bibinfo {pages} {034309} (\bibinfo {year} {2022})}\BibitemShut
  {NoStop}%
\bibitem [{\citenamefont {Jokiniemi}\ \emph {et~al.}(2023)\citenamefont
  {Jokiniemi}, \citenamefont {Romeo}, \citenamefont {Soriano},\ and\
  \citenamefont {Men\'endez}}]{Jokiniemi:2023}%
  \BibitemOpen
  \bibfield  {author} {\bibinfo {author} {\bibfnamefont {L.}~\bibnamefont
  {Jokiniemi}}, \bibinfo {author} {\bibfnamefont {B.}~\bibnamefont {Romeo}},
  \bibinfo {author} {\bibfnamefont {P.}~\bibnamefont {Soriano}},\ and\ \bibinfo
  {author} {\bibfnamefont {J.}~\bibnamefont {Men\'endez}},\ }\bibfield  {title}
  {\bibinfo {title} {{Neutrinoless \ensuremath{\beta}\ensuremath{\beta}-decay
  nuclear matrix elements from two-neutrino
  \ensuremath{\beta}\ensuremath{\beta}-decay data}},\ }\href
  {https://doi.org/10.1103/PhysRevC.107.044305} {\bibfield  {journal} {\bibinfo
   {journal} {Phys. Rev. C}\ }\textbf {\bibinfo {volume} {107}},\ \bibinfo
  {pages} {044305} (\bibinfo {year} {2023})},\ \Eprint
  {https://arxiv.org/abs/2207.05108} {arXiv:2207.05108 [nucl-th]} \BibitemShut
  {NoStop}%
\bibitem [{Sup()}]{SupplementalMaterial}%
  \BibitemOpen
  \bibfield  {title} {\bibinfo {title} {See supplemental material},\ }\href
  {https://link.aps.org/supplemental/} {\ }\BibitemShut {NoStop}%
\bibitem [{\citenamefont {Jiang}\ \emph {et~al.}(2020)\citenamefont {Jiang},
  \citenamefont {Ekstr\"om}, \citenamefont {Forss\'en}, \citenamefont {Hagen},
  \citenamefont {Jansen},\ and\ \citenamefont {Papenbrock}}]{Jiang:2020}%
  \BibitemOpen
  \bibfield  {author} {\bibinfo {author} {\bibfnamefont {W.~G.}\ \bibnamefont
  {Jiang}}, \bibinfo {author} {\bibfnamefont {A.}~\bibnamefont {Ekstr\"om}},
  \bibinfo {author} {\bibfnamefont {C.}~\bibnamefont {Forss\'en}}, \bibinfo
  {author} {\bibfnamefont {G.}~\bibnamefont {Hagen}}, \bibinfo {author}
  {\bibfnamefont {G.~R.}\ \bibnamefont {Jansen}},\ and\ \bibinfo {author}
  {\bibfnamefont {T.}~\bibnamefont {Papenbrock}},\ }\bibfield  {title}
  {\bibinfo {title} {{Accurate bulk properties of nuclei from $A=2$ to $\infty$
  from potentials with $\Delta$ isobars}},\ }\href
  {https://doi.org/10.1103/PhysRevC.102.054301} {\bibfield  {journal} {\bibinfo
   {journal} {Phys. Rev. C}\ }\textbf {\bibinfo {volume} {102}},\ \bibinfo
  {pages} {054301} (\bibinfo {year} {2020})}\BibitemShut {NoStop}%
\bibitem [{\citenamefont {Melendez}\ \emph {et~al.}(2019)\citenamefont
  {Melendez}, \citenamefont {Furnstahl}, \citenamefont {Phillips},
  \citenamefont {Pratola},\ and\ \citenamefont {Wesolowski}}]{Melendez2019}%
  \BibitemOpen
  \bibfield  {author} {\bibinfo {author} {\bibfnamefont {J.~A.}\ \bibnamefont
  {Melendez}}, \bibinfo {author} {\bibfnamefont {R.~J.}\ \bibnamefont
  {Furnstahl}}, \bibinfo {author} {\bibfnamefont {D.~R.}\ \bibnamefont
  {Phillips}}, \bibinfo {author} {\bibfnamefont {M.~T.}\ \bibnamefont
  {Pratola}},\ and\ \bibinfo {author} {\bibfnamefont {S.}~\bibnamefont
  {Wesolowski}},\ }\bibfield  {title} {\bibinfo {title} {Quantifying correlated
  truncation errors in effective field theory},\ }\href
  {https://doi.org/10.1103/PhysRevC.100.044001} {\bibfield  {journal} {\bibinfo
   {journal} {Phys. Rev. C}\ }\textbf {\bibinfo {volume} {100}},\ \bibinfo
  {pages} {044001} (\bibinfo {year} {2019})}\BibitemShut {NoStop}%
\bibitem [{\citenamefont {Entem}\ \emph {et~al.}(2017)\citenamefont {Entem},
  \citenamefont {Machleidt},\ and\ \citenamefont {Nosyk}}]{Entem2017}%
  \BibitemOpen
  \bibfield  {author} {\bibinfo {author} {\bibfnamefont {D.~R.}\ \bibnamefont
  {Entem}}, \bibinfo {author} {\bibfnamefont {R.}~\bibnamefont {Machleidt}},\
  and\ \bibinfo {author} {\bibfnamefont {Y.}~\bibnamefont {Nosyk}},\ }\bibfield
   {title} {\bibinfo {title} {High-quality two-nucleon potentials up to fifth
  order of the chiral expansion},\ }\href
  {https://doi.org/10.1103/PhysRevC.96.024004} {\bibfield  {journal} {\bibinfo
  {journal} {Phys. Rev. C}\ }\textbf {\bibinfo {volume} {96}},\ \bibinfo
  {pages} {024004} (\bibinfo {year} {2017})}\BibitemShut {NoStop}%
\bibitem [{\citenamefont {Kravvaris}\ \emph {et~al.}(2020)\citenamefont
  {Kravvaris}, \citenamefont {Quinlan}, \citenamefont {Quaglioni},
  \citenamefont {Wendt},\ and\ \citenamefont
  {Navr\'atil}}]{Konstantinos:2020PRC}%
  \BibitemOpen
  \bibfield  {author} {\bibinfo {author} {\bibfnamefont {K.}~\bibnamefont
  {Kravvaris}}, \bibinfo {author} {\bibfnamefont {K.~R.}\ \bibnamefont
  {Quinlan}}, \bibinfo {author} {\bibfnamefont {S.}~\bibnamefont {Quaglioni}},
  \bibinfo {author} {\bibfnamefont {K.~A.}\ \bibnamefont {Wendt}},\ and\
  \bibinfo {author} {\bibfnamefont {P.}~\bibnamefont {Navr\'atil}},\ }\bibfield
   {title} {\bibinfo {title} {Quantifying uncertainties in
  neutron-$\ensuremath{\alpha}$ scattering with chiral nucleon-nucleon and
  three-nucleon forces},\ }\href {https://doi.org/10.1103/PhysRevC.102.024616}
  {\bibfield  {journal} {\bibinfo  {journal} {Phys. Rev. C}\ }\textbf {\bibinfo
  {volume} {102}},\ \bibinfo {pages} {024616} (\bibinfo {year}
  {2020})}\BibitemShut {NoStop}%
\bibitem [{\citenamefont {Stroberg}\ \emph {et~al.}(2019)\citenamefont
  {Stroberg}, \citenamefont {Hergert}, \citenamefont {Bogner},\ and\
  \citenamefont {Holt}}]{Stroberg:2019}%
  \BibitemOpen
  \bibfield  {author} {\bibinfo {author} {\bibfnamefont {S.~R.}\ \bibnamefont
  {Stroberg}}, \bibinfo {author} {\bibfnamefont {H.}~\bibnamefont {Hergert}},
  \bibinfo {author} {\bibfnamefont {S.~K.}\ \bibnamefont {Bogner}},\ and\
  \bibinfo {author} {\bibfnamefont {J.~D.}\ \bibnamefont {Holt}},\ }\bibfield
  {title} {\bibinfo {title} {Nonempirical interactions for the nuclear shell
  model: An update},\ }\href
  {https://doi.org/10.1146/annurev-nucl-101917-021120} {\bibfield  {journal}
  {\bibinfo  {journal} {Annu. Rev. Nucl. Part. Sci.}\ }\textbf {\bibinfo
  {volume} {69}},\ \bibinfo {pages} {307} (\bibinfo {year} {2019})}\BibitemShut
  {NoStop}%
\bibitem [{\citenamefont {Simonis}\ \emph {et~al.}(2017)\citenamefont
  {Simonis}, \citenamefont {Stroberg}, \citenamefont {Hebeler}, \citenamefont
  {Holt},\ and\ \citenamefont {Schwenk}}]{Simonis:2017}%
  \BibitemOpen
  \bibfield  {author} {\bibinfo {author} {\bibfnamefont {J.}~\bibnamefont
  {Simonis}}, \bibinfo {author} {\bibfnamefont {S.~R.}\ \bibnamefont
  {Stroberg}}, \bibinfo {author} {\bibfnamefont {K.}~\bibnamefont {Hebeler}},
  \bibinfo {author} {\bibfnamefont {J.~D.}\ \bibnamefont {Holt}},\ and\
  \bibinfo {author} {\bibfnamefont {A.}~\bibnamefont {Schwenk}},\ }\bibfield
  {title} {\bibinfo {title} {Saturation with chiral interactions and
  consequences for finite nuclei},\ }\href
  {https://doi.org/10.1103/PhysRevC.96.014303} {\bibfield  {journal} {\bibinfo
  {journal} {Phys. Rev. C}\ }\textbf {\bibinfo {volume} {96}},\ \bibinfo
  {pages} {014303} (\bibinfo {year} {2017})}\BibitemShut {NoStop}%
\bibitem [{\citenamefont {Ayangeakaa}\ \emph {et~al.}(2023)\citenamefont
  {Ayangeakaa}, \citenamefont {Janssens}, \citenamefont {Zhu}, \citenamefont
  {Allmond}, \citenamefont {Brown}, \citenamefont {Wu}, \citenamefont {Albers},
  \citenamefont {Auranen}, \citenamefont {Bucher}, \citenamefont {Carpenter},
  \citenamefont {Chowdhury}, \citenamefont {Cline}, \citenamefont {Crawford},
  \citenamefont {Fallon}, \citenamefont {Forney}, \citenamefont {Gade},
  \citenamefont {Hartley}, \citenamefont {Hayes}, \citenamefont {Henderson},
  \citenamefont {Kondev}, \citenamefont {Krishichayan}, \citenamefont
  {Lauritsen}, \citenamefont {Li}, \citenamefont {Little}, \citenamefont
  {Macchiavelli}, \citenamefont {Rhodes}, \citenamefont {Seweryniak},
  \citenamefont {Stolze}, \citenamefont {Walters},\ and\ \citenamefont
  {Wu}}]{Ayangeakaa:2023PRC}%
  \BibitemOpen
  \bibfield  {author} {\bibinfo {author} {\bibfnamefont {A.~D.}\ \bibnamefont
  {Ayangeakaa}}, \bibinfo {author} {\bibfnamefont {R.~V.~F.}\ \bibnamefont
  {Janssens}}, \bibinfo {author} {\bibfnamefont {S.}~\bibnamefont {Zhu}},
  \bibinfo {author} {\bibfnamefont {J.~M.}\ \bibnamefont {Allmond}}, \bibinfo
  {author} {\bibfnamefont {B.~A.}\ \bibnamefont {Brown}}, \bibinfo {author}
  {\bibfnamefont {C.~Y.}\ \bibnamefont {Wu}}, \bibinfo {author} {\bibfnamefont
  {M.}~\bibnamefont {Albers}}, \bibinfo {author} {\bibfnamefont
  {K.}~\bibnamefont {Auranen}}, \bibinfo {author} {\bibfnamefont
  {B.}~\bibnamefont {Bucher}}, \bibinfo {author} {\bibfnamefont {M.~P.}\
  \bibnamefont {Carpenter}}, \bibinfo {author} {\bibfnamefont {P.}~\bibnamefont
  {Chowdhury}}, \bibinfo {author} {\bibfnamefont {D.}~\bibnamefont {Cline}},
  \bibinfo {author} {\bibfnamefont {H.~L.}\ \bibnamefont {Crawford}}, \bibinfo
  {author} {\bibfnamefont {P.}~\bibnamefont {Fallon}}, \bibinfo {author}
  {\bibfnamefont {A.~M.}\ \bibnamefont {Forney}}, \bibinfo {author}
  {\bibfnamefont {A.}~\bibnamefont {Gade}}, \bibinfo {author} {\bibfnamefont
  {D.~J.}\ \bibnamefont {Hartley}}, \bibinfo {author} {\bibfnamefont {A.~B.}\
  \bibnamefont {Hayes}}, \bibinfo {author} {\bibfnamefont {J.}~\bibnamefont
  {Henderson}}, \bibinfo {author} {\bibfnamefont {F.~G.}\ \bibnamefont
  {Kondev}}, \bibinfo {author} {\bibnamefont {Krishichayan}}, \bibinfo {author}
  {\bibfnamefont {T.}~\bibnamefont {Lauritsen}}, \bibinfo {author}
  {\bibfnamefont {J.}~\bibnamefont {Li}}, \bibinfo {author} {\bibfnamefont
  {D.}~\bibnamefont {Little}}, \bibinfo {author} {\bibfnamefont {A.~O.}\
  \bibnamefont {Macchiavelli}}, \bibinfo {author} {\bibfnamefont
  {D.}~\bibnamefont {Rhodes}}, \bibinfo {author} {\bibfnamefont
  {D.}~\bibnamefont {Seweryniak}}, \bibinfo {author} {\bibfnamefont {S.~M.}\
  \bibnamefont {Stolze}}, \bibinfo {author} {\bibfnamefont {W.~B.}\
  \bibnamefont {Walters}},\ and\ \bibinfo {author} {\bibfnamefont
  {J.}~\bibnamefont {Wu}},\ }\bibfield  {title} {\bibinfo {title} {Triaxiality
  and the nature of low-energy excitations in $^{76}\mathrm{Ge}$},\ }\href
  {https://doi.org/10.1103/PhysRevC.107.044314} {\bibfield  {journal} {\bibinfo
   {journal} {Phys. Rev. C}\ }\textbf {\bibinfo {volume} {107}},\ \bibinfo
  {pages} {044314} (\bibinfo {year} {2023})}\BibitemShut {NoStop}%
\bibitem [{\citenamefont {Henderson}\ \emph {et~al.}(2019)\citenamefont
  {Henderson}, \citenamefont {Wu}, \citenamefont {Ash}, \citenamefont {Brown},
  \citenamefont {Bender}, \citenamefont {Elder}, \citenamefont {Elman},
  \citenamefont {Gade}, \citenamefont {Grinder}, \citenamefont {Iwasaki},
  \citenamefont {Longfellow}, \citenamefont {Mijatovi\ifmmode~\acute{c}\else
  \'{c}\fi{}}, \citenamefont {Rhodes}, \citenamefont {Spieker},\ and\
  \citenamefont {Weisshaar}}]{Henderson:2019}%
  \BibitemOpen
  \bibfield  {author} {\bibinfo {author} {\bibfnamefont {J.}~\bibnamefont
  {Henderson}}, \bibinfo {author} {\bibfnamefont {C.~Y.}\ \bibnamefont {Wu}},
  \bibinfo {author} {\bibfnamefont {J.}~\bibnamefont {Ash}}, \bibinfo {author}
  {\bibfnamefont {B.~A.}\ \bibnamefont {Brown}}, \bibinfo {author}
  {\bibfnamefont {P.~C.}\ \bibnamefont {Bender}}, \bibinfo {author}
  {\bibfnamefont {R.}~\bibnamefont {Elder}}, \bibinfo {author} {\bibfnamefont
  {B.}~\bibnamefont {Elman}}, \bibinfo {author} {\bibfnamefont
  {A.}~\bibnamefont {Gade}}, \bibinfo {author} {\bibfnamefont {M.}~\bibnamefont
  {Grinder}}, \bibinfo {author} {\bibfnamefont {H.}~\bibnamefont {Iwasaki}},
  \bibinfo {author} {\bibfnamefont {B.}~\bibnamefont {Longfellow}}, \bibinfo
  {author} {\bibfnamefont {T.}~\bibnamefont {Mijatovi\ifmmode~\acute{c}\else
  \'{c}\fi{}}}, \bibinfo {author} {\bibfnamefont {D.}~\bibnamefont {Rhodes}},
  \bibinfo {author} {\bibfnamefont {M.}~\bibnamefont {Spieker}},\ and\ \bibinfo
  {author} {\bibfnamefont {D.}~\bibnamefont {Weisshaar}},\ }\bibfield  {title}
  {\bibinfo {title} {Triaxiality in selenium-76},\ }\href
  {https://doi.org/10.1103/PhysRevC.99.054313} {\bibfield  {journal} {\bibinfo
  {journal} {Phys. Rev. C}\ }\textbf {\bibinfo {volume} {99}},\ \bibinfo
  {pages} {054313} (\bibinfo {year} {2019})}\BibitemShut {NoStop}%
\bibitem [{\citenamefont {Rodr\'\i{}guez}(2017)}]{Rodriguez:2017JPG}%
  \BibitemOpen
  \bibfield  {author} {\bibinfo {author} {\bibfnamefont {T.}~\bibnamefont
  {Rodr\'\i{}guez}},\ }\bibfield  {title} {\bibinfo {title} {{Role of
  triaxiality in $^{76}\mathrm{Ge}$ and $^{76}\mathrm{Se}$ nuclei studied with
  Gogny energy density functionals}},\ }\href
  {https://doi.org/10.1088/1361-6471/aa57d3} {\bibfield  {journal} {\bibinfo
  {journal} {J. Phys. G}\ }\textbf {\bibinfo {volume} {44}},\ \bibinfo {pages}
  {034002} (\bibinfo {year} {2017})}\BibitemShut {NoStop}%
\bibitem [{\citenamefont {Henderson}\ \emph {et~al.}(2018)\citenamefont
  {Henderson} \emph {et~al.}}]{Hend18E2}%
  \BibitemOpen
  \bibfield  {author} {\bibinfo {author} {\bibfnamefont {J.}~\bibnamefont
  {Henderson}} \emph {et~al.},\ }\bibfield  {title} {\bibinfo {title} {{Testing
  microscopically derived descriptions of nuclear collectivity: Coulomb
  excitation of $^{22}$Mg}},\ }\href
  {https://doi.org/10.1016/j.physletb.2018.05.064} {\bibfield  {journal}
  {\bibinfo  {journal} {Phys. Lett. B}\ }\textbf {\bibinfo {volume} {782}},\
  \bibinfo {pages} {468} (\bibinfo {year} {2018})}\BibitemShut {NoStop}%
\bibitem [{\citenamefont {Stroberg}\ \emph {et~al.}(2022)\citenamefont
  {Stroberg}, \citenamefont {Henderson}, \citenamefont {Hackman}, \citenamefont
  {Ruotsalainen}, \citenamefont {Hagen},\ and\ \citenamefont
  {Holt}}]{Stro22E2}%
  \BibitemOpen
  \bibfield  {author} {\bibinfo {author} {\bibfnamefont {S.~R.}\ \bibnamefont
  {Stroberg}}, \bibinfo {author} {\bibfnamefont {J.}~\bibnamefont {Henderson}},
  \bibinfo {author} {\bibfnamefont {G.}~\bibnamefont {Hackman}}, \bibinfo
  {author} {\bibfnamefont {P.}~\bibnamefont {Ruotsalainen}}, \bibinfo {author}
  {\bibfnamefont {G.}~\bibnamefont {Hagen}},\ and\ \bibinfo {author}
  {\bibfnamefont {J.~D.}\ \bibnamefont {Holt}},\ }\bibfield  {title} {\bibinfo
  {title} {{Systematics of E2 strength in the sd shell with the valence-space
  in-medium similarity renormalization group}},\ }\href
  {https://doi.org/10.1103/PhysRevC.105.034333} {\bibfield  {journal} {\bibinfo
   {journal} {Phys. Rev. C}\ }\textbf {\bibinfo {volume} {105}},\ \bibinfo
  {pages} {034333} (\bibinfo {year} {2022})}\BibitemShut {NoStop}%
\bibitem [{sup()}]{supp}%
  \BibitemOpen
  \bibfield  {title} {\bibinfo {title} {See supplemental material},\
  }\href@noop {} {\ }\BibitemShut {NoStop}%
\bibitem [{\citenamefont {Sen'kov}\ and\ \citenamefont
  {Horoi}(2013)}]{Senkov2013}%
  \BibitemOpen
  \bibfield  {author} {\bibinfo {author} {\bibfnamefont {R.~A.}\ \bibnamefont
  {Sen'kov}}\ and\ \bibinfo {author} {\bibfnamefont {M.}~\bibnamefont
  {Horoi}},\ }\bibfield  {title} {\bibinfo {title} {Neutrinoless
  double-$\ensuremath{\beta}$ decay of $^{48}\mathrm{Ca}$ in the shell model:
  Closure versus nonclosure approximation},\ }\href
  {https://doi.org/10.1103/PhysRevC.88.064312} {\bibfield  {journal} {\bibinfo
  {journal} {Phys. Rev. C}\ }\textbf {\bibinfo {volume} {88}},\ \bibinfo
  {pages} {064312} (\bibinfo {year} {2013})}\BibitemShut {NoStop}%
\bibitem [{\citenamefont {Wang}\ \emph {et~al.}(2021)\citenamefont {Wang},
  \citenamefont {Zhao},\ and\ \citenamefont {Meng}}]{Wang:2021}%
  \BibitemOpen
  \bibfield  {author} {\bibinfo {author} {\bibfnamefont {Y.~K.}\ \bibnamefont
  {Wang}}, \bibinfo {author} {\bibfnamefont {P.~W.}\ \bibnamefont {Zhao}},\
  and\ \bibinfo {author} {\bibfnamefont {J.}~\bibnamefont {Meng}},\ }\bibfield
  {title} {\bibinfo {title} {Nuclear matrix elements of neutrinoless
  double-$\ensuremath{\beta}$ decay in the triaxial projected shell model},\
  }\href {https://doi.org/10.1103/PhysRevC.104.014320} {\bibfield  {journal}
  {\bibinfo  {journal} {Phys. Rev. C}\ }\textbf {\bibinfo {volume} {104}},\
  \bibinfo {pages} {014320} (\bibinfo {year} {2021})}\BibitemShut {NoStop}%
\bibitem [{\citenamefont {Cirigliano}\ \emph
  {et~al.}(2018{\natexlab{b}})\citenamefont {Cirigliano}, \citenamefont
  {Dekens}, \citenamefont {Mereghetti},\ and\ \citenamefont
  {Walker-Loud}}]{Cirigliano:2018PRC}%
  \BibitemOpen
  \bibfield  {author} {\bibinfo {author} {\bibfnamefont {V.}~\bibnamefont
  {Cirigliano}}, \bibinfo {author} {\bibfnamefont {W.}~\bibnamefont {Dekens}},
  \bibinfo {author} {\bibfnamefont {E.}~\bibnamefont {Mereghetti}},\ and\
  \bibinfo {author} {\bibfnamefont {A.}~\bibnamefont {Walker-Loud}},\
  }\bibfield  {title} {\bibinfo {title} {Neutrinoless
  double-$\ensuremath{\beta}$ decay in effective field theory: The
  light-majorana neutrino-exchange mechanism},\ }\href
  {https://doi.org/10.1103/PhysRevC.97.065501} {\bibfield  {journal} {\bibinfo
  {journal} {Phys. Rev. C}\ }\textbf {\bibinfo {volume} {97}},\ \bibinfo
  {pages} {065501} (\bibinfo {year} {2018}{\natexlab{b}})}\BibitemShut
  {NoStop}%
\bibitem [{\citenamefont {Wirth}\ \emph {et~al.}(2021)\citenamefont {Wirth},
  \citenamefont {Yao},\ and\ \citenamefont {Hergert}}]{Wirth:2021}%
  \BibitemOpen
  \bibfield  {author} {\bibinfo {author} {\bibfnamefont {R.}~\bibnamefont
  {Wirth}}, \bibinfo {author} {\bibfnamefont {J.~M.}\ \bibnamefont {Yao}},\
  and\ \bibinfo {author} {\bibfnamefont {H.}~\bibnamefont {Hergert}},\
  }\bibfield  {title} {\bibinfo {title} {Ab initio calculation of the contact
  operator contribution in the standard mechanism for neutrinoless double beta
  decay},\ }\href {https://doi.org/10.1103/PhysRevLett.127.242502} {\bibfield
  {journal} {\bibinfo  {journal} {Phys. Rev. Lett.}\ }\textbf {\bibinfo
  {volume} {127}},\ \bibinfo {pages} {242502} (\bibinfo {year}
  {2021})}\BibitemShut {NoStop}%
\bibitem [{\citenamefont {Cirigliano}\ \emph {et~al.}(2021)\citenamefont
  {Cirigliano}, \citenamefont {Dekens}, \citenamefont {de~Vries}, \citenamefont
  {Hoferichter},\ and\ \citenamefont {Mereghetti}}]{Cirigliano:2021PRL}%
  \BibitemOpen
  \bibfield  {author} {\bibinfo {author} {\bibfnamefont {V.}~\bibnamefont
  {Cirigliano}}, \bibinfo {author} {\bibfnamefont {W.}~\bibnamefont {Dekens}},
  \bibinfo {author} {\bibfnamefont {J.}~\bibnamefont {de~Vries}}, \bibinfo
  {author} {\bibfnamefont {M.}~\bibnamefont {Hoferichter}},\ and\ \bibinfo
  {author} {\bibfnamefont {E.}~\bibnamefont {Mereghetti}},\ }\bibfield  {title}
  {\bibinfo {title} {Toward complete leading-order predictions for neutrinoless
  double $\ensuremath{\beta}$ decay},\ }\href
  {https://doi.org/10.1103/PhysRevLett.126.172002} {\bibfield  {journal}
  {\bibinfo  {journal} {Phys. Rev. Lett.}\ }\textbf {\bibinfo {volume} {126}},\
  \bibinfo {pages} {172002} (\bibinfo {year} {2021})}\BibitemShut {NoStop}%
\bibitem [{\citenamefont {Pastore}\ \emph {et~al.}(2018)\citenamefont
  {Pastore}, \citenamefont {Carlson}, \citenamefont {Cirigliano}, \citenamefont
  {Dekens}, \citenamefont {Mereghetti},\ and\ \citenamefont
  {Wiringa}}]{Pastore:2018}%
  \BibitemOpen
  \bibfield  {author} {\bibinfo {author} {\bibfnamefont {S.}~\bibnamefont
  {Pastore}}, \bibinfo {author} {\bibfnamefont {J.}~\bibnamefont {Carlson}},
  \bibinfo {author} {\bibfnamefont {V.}~\bibnamefont {Cirigliano}}, \bibinfo
  {author} {\bibfnamefont {W.}~\bibnamefont {Dekens}}, \bibinfo {author}
  {\bibfnamefont {E.}~\bibnamefont {Mereghetti}},\ and\ \bibinfo {author}
  {\bibfnamefont {R.~B.}\ \bibnamefont {Wiringa}},\ }\bibfield  {title}
  {\bibinfo {title} {Neutrinoless double-$\ensuremath{\beta}$ decay matrix
  elements in light nuclei},\ }\href
  {https://doi.org/10.1103/PhysRevC.97.014606} {\bibfield  {journal} {\bibinfo
  {journal} {Phys. Rev. C}\ }\textbf {\bibinfo {volume} {97}},\ \bibinfo
  {pages} {014606} (\bibinfo {year} {2018})}\BibitemShut {NoStop}%
\bibitem [{\citenamefont {Som\`a}\ \emph {et~al.}(2020)\citenamefont {Som\`a},
  \citenamefont {Navr\'atil}, \citenamefont {Raimondi}, \citenamefont
  {Barbieri},\ and\ \citenamefont {Duguet}}]{Soma:2020}%
  \BibitemOpen
  \bibfield  {author} {\bibinfo {author} {\bibfnamefont {V.}~\bibnamefont
  {Som\`a}}, \bibinfo {author} {\bibfnamefont {P.}~\bibnamefont {Navr\'atil}},
  \bibinfo {author} {\bibfnamefont {F.}~\bibnamefont {Raimondi}}, \bibinfo
  {author} {\bibfnamefont {C.}~\bibnamefont {Barbieri}},\ and\ \bibinfo
  {author} {\bibfnamefont {T.}~\bibnamefont {Duguet}},\ }\bibfield  {title}
  {\bibinfo {title} {{Novel chiral Hamiltonian and observables in light and
  medium-mass nuclei}},\ }\href {https://doi.org/10.1103/PhysRevC.101.014318}
  {\bibfield  {journal} {\bibinfo  {journal} {Phys. Rev. C}\ }\textbf {\bibinfo
  {volume} {101}},\ \bibinfo {pages} {014318} (\bibinfo {year} {2020})},\
  \Eprint {https://arxiv.org/abs/1907.09790} {arXiv:1907.09790 [nucl-th]}
  \BibitemShut {NoStop}%
\bibitem [{\citenamefont {Kotila}\ and\ \citenamefont
  {Iachello}(2012)}]{Kotila:2012}%
  \BibitemOpen
  \bibfield  {author} {\bibinfo {author} {\bibfnamefont {J.}~\bibnamefont
  {Kotila}}\ and\ \bibinfo {author} {\bibfnamefont {F.}~\bibnamefont
  {Iachello}},\ }\bibfield  {title} {\bibinfo {title} {Phase-space factors for
  double-$\ensuremath{\beta}$ decay},\ }\href
  {https://doi.org/10.1103/PhysRevC.85.034316} {\bibfield  {journal} {\bibinfo
  {journal} {Phys. Rev. C}\ }\textbf {\bibinfo {volume} {85}},\ \bibinfo
  {pages} {034316} (\bibinfo {year} {2012})}\BibitemShut {NoStop}%
\bibitem [{\citenamefont {Stef\'anik}\ \emph {et~al.}(2015)\citenamefont
  {Stef\'anik}, \citenamefont {Dvornick\'y}, \citenamefont
  {\ifmmode~\check{S}\else \v{S}\fi{}imkovic},\ and\ \citenamefont
  {Vogel}}]{Stefanik:2015}%
  \BibitemOpen
  \bibfield  {author} {\bibinfo {author} {\bibfnamefont {D.}~\bibnamefont
  {Stef\'anik}}, \bibinfo {author} {\bibfnamefont {R.}~\bibnamefont
  {Dvornick\'y}}, \bibinfo {author} {\bibfnamefont {F.}~\bibnamefont
  {\ifmmode~\check{S}\else \v{S}\fi{}imkovic}},\ and\ \bibinfo {author}
  {\bibfnamefont {P.}~\bibnamefont {Vogel}},\ }\bibfield  {title} {\bibinfo
  {title} {Reexamining the light neutrino exchange mechanism of the
  $0\ensuremath{\nu}\ensuremath{\beta}\ensuremath{\beta}$ decay with left- and
  right-handed leptonic and hadronic currents},\ }\href
  {https://doi.org/10.1103/PhysRevC.92.055502} {\bibfield  {journal} {\bibinfo
  {journal} {Phys. Rev. C}\ }\textbf {\bibinfo {volume} {92}},\ \bibinfo
  {pages} {055502} (\bibinfo {year} {2015})}\BibitemShut {NoStop}%
\bibitem [{\citenamefont {Abgrall}\ \emph {et~al.}(2021)\citenamefont {Abgrall}
  \emph {et~al.}}]{LEGEND:Abgrall-2021}%
  \BibitemOpen
  \bibfield  {author} {\bibinfo {author} {\bibfnamefont {N.}~\bibnamefont
  {Abgrall}} \emph {et~al.} (\bibinfo {collaboration} {LEGEND}),\ }\bibfield
  {title} {\bibinfo {title} {{The Large Enriched Germanium Experiment for
  Neutrinoless Double Beta Decay (LEGEND)}},\ }\href
  {https://doi.org/10.1063/1.5007652} {\bibfield  {journal} {\bibinfo
  {journal} {AIP Conf. Proc.}\ }\textbf {\bibinfo {volume} {1894}},\ \bibinfo
  {pages} {020027} (\bibinfo {year} {2021})}\BibitemShut {NoStop}%
\bibitem [{\citenamefont {Belley}\ \emph {et~al.}(2023)\citenamefont {Belley},
  \citenamefont {Miyagi}, \citenamefont {Stroberg},\ and\ \citenamefont
  {Holt}}]{Belley2023TeXe}%
  \BibitemOpen
  \bibfield  {author} {\bibinfo {author} {\bibfnamefont {A.}~\bibnamefont
  {Belley}}, \bibinfo {author} {\bibfnamefont {T.}~\bibnamefont {Miyagi}},
  \bibinfo {author} {\bibfnamefont {S.~R.}\ \bibnamefont {Stroberg}},\ and\
  \bibinfo {author} {\bibfnamefont {J.~D.}\ \bibnamefont {Holt}},\ }\href@noop
  {} {\bibinfo {title} {Ab initio calculations of neutrinoless $\beta \beta$
  decay refine neutrino mass limits}} (\bibinfo {year} {2023}),\ \Eprint
  {https://arxiv.org/abs/arXiv:2307.15156} {arXiv:arXiv:2307.15156 [nucl-th]}
  \BibitemShut {NoStop}%
\end{thebibliography}


\end{document}